\documentclass[prd,aps,eqsecnum,floatfix,nofootinbib,preprint,tightenlines]{revtex4}

\usepackage{latexsym}
\usepackage{graphicx}
\usepackage{multirow}
\usepackage{color}
\usepackage{amsmath}
\usepackage{hyperref}  

\def\floatcaption#1#2{ \caption{#2 \label{#1}} }

\def\bibi{\bibitem}

\def\b{\beta}
\def\c{\chi}
\def\d{\delta}
\def\e{\epsilon}                
\def\g{\gamma}

\def\i{\iota}
\def\j{\psi}

\def\l{\lambda}
\def\m{\mu}
\def\n{\nu}
\def\o{\omega}
\def\p{\pi}                     
\def\r{\rho}                    
\def\s{\sigma}                  

\def\F{\Phi}

\def\J{\Psi}
\def\L{\Lambda}
\def\O{\Omega}
\def\P{\Pi}

\def\S{\Sigma}
\def\U{\Upsilon}

\def\cl{{\cal L}}

\def\cbo{{\,\raise-.15ex\Sc [\,}}                       

\def\sl#1{\rlap{\hbox{$\mskip 1 mu /$}}#1}      

\def\svev#1{\left\langle #1\right\rangle}       

\def\ddt#1{{\buildrel {\hbox{\LARGE .\kern-2pt.}} \over {#1}}}

\def\ie{\mbox{\it i.e.}}
\def\eg{\mbox{\it e.g.}}

\def\geqx{\,\raisebox{-1.0ex}{$\stackrel{\textstyle >}{\sim}$}\,}

\def\tr{{\rm tr}\,}

\def\half{{1\over 2}}

\def\ttl#1{{\it #1}}

\long \def \blockcomment #1\endcomment{}

\def\seef{{\it cf.\  }}

\def\tr{\,{\rm tr}}

\def\tB{\tilde{B}}

\def\btB{\overline{\tB}}
\def\hv{\hat{v}}
\def\hbv{\hat{\overline{v}}}
\def\hT{\hat{T}}

\def\bB{\overline{B}}
\def\bU{\bar{\U}}
\def\bc{\overline{\c}}
\def\bj{\overline{\j}}
\def\bJ{\bar{\J}}
\def\bT{\overline{T}}
\def\bt{\overline{t}}
\def\bv{\overline{v}}
\def\tom{\tilde\o}
\def\bto{\bar{\tilde\o}}

\def\bom{\overline{\o}}

\def\tJ{\tilde\J}
\def\btJ{\bar{\tilde\J}}
\def\tU{\tilde\U}
\def\btU{\bar{\tilde\U}}

\def\one{\mbox{\bf 1}}
\def\three{\mbox{\bf 3}}
\def\threebar{\mbox{${\bf \overline{3}}$}}
\def\five{\mbox{\bf 5}}
\def\fivebar{\mbox{${\bf \overline{5}}$}}

\def\irrep{{\it irrep}}
\def\irreps{{\it irreps}}

\begin{document}

\begin{center}
{\large\bf Top quark induced effective potential in a composite Higgs model}\\[8mm]
Maarten Golterman$^a$ and Yigal Shamir$^b$\\[10 mm]
{\small
$^a$Department of Physics and Astronomy, San Francisco State University,\\
San Francisco, CA 94132, USA\\
$^b$Raymond and Beverly Sackler School of Physics and Astronomy,\\
Tel~Aviv University, 69978, Tel~Aviv, Israel}\\[10mm]
\end{center}

\begin{quotation}
We consider non-perturbative aspects of a composite Higgs model
that serves as a prototype for physics beyond the Standard Model, in which
a new strongly interacting sector undergoes chiral symmetry breaking,
and generates the Higgs particle as a pseudo Nambu--Goldstone boson.
In addition, the top quark couples linearly to baryons of the new strong
sector, thereby becoming partially composite.
We study the dynamics leading to the top quark Yukawa coupling
as well as the top quark contribution
to the effective potential for the Higgs,
obtaining expressions for these couplings
in terms of baryonic correlation functions in the underlying
strongly interacting theory.   We then
show that a large-$N$ limit exists in which the top quark contribution to the
Higgs effective potential overcomes that of the weak gauge bosons,
inducing electroweak symmetry breaking.  The same large-$N$ limit also suggests
that the baryons that couple to the top quark may be relatively light.
This composite Higgs model, and similar ones, can be studied on the
lattice with the methods developed for lattice QCD.
\end{quotation}

\newpage
\section{\label{introduction} Introduction}
Since the discovery of a light Higgs boson at the LHC,
interest in beyond-the-Standard-Model scenarios
has focused on models in which the Higgs is naturally light
compared to the typical scale of new physics.
One approach postulates the existence of a new strongly interacting sector,
which we will refer to as hypercolor in this paper.  The Higgs doublet of the Standard Model (SM)
emerges among the Nambu--Goldstone bosons (NGBs)
originating from dynamical symmetry breaking of the flavor symmetry group $G$
of the hypercolor theory.
The electroweak gauge bosons as well as the SM fermions then couple to these
NGBs, breaking the symmetry group $G$ explicitly to a smaller group,
thereby generating an effective potential for the NGBs.
Under suitable conditions, this radiatively induced effective potential
leads to electroweak symmetry breaking, with
the Higgs field acquiring an expectation value as in the SM.
This framework still allows for many different possibilities.
For reviews that span the evolution of this field, as well as for
generic features of these models, we refer to Refs.~\cite{MP,GGPR,BBRV,RC,BCS}.

We will be interested in composite-Higgs models in which
the sector external to the hypercolor gauge theory,
which includes the SM gauge bosons and fermions, is as simple as possible.
For instance, we do not wish to introduce any weakly coupled gauge
bosons besides the electroweak gauge bosons,
as in Little Higgs models \cite {MP}.
The electroweak gauge bosons have to stay massless at the
dynamical symmetry breaking scale of the hypercolor theory,
and therefore they have to couple to generators in the unbroken
flavor subgroup $H\subset G$.   As a result,
the effective potential generated for the
hypercolor NGBs by the electroweak gauge bosons will not lead to
electroweak symmetry breaking, a phenomenon often referred to as
vacuum alignment \cite{vac}.

Electroweak symmetry breaking must therefore originate
in the effective potential generated by the top quark,
being the SM fermion with the strongest coupling to the Higgs,
and, hence, to the hypercolor theory.  We will postulate that the top quark
couples linearly to hyperbaryons (the baryons of the hypercolor theory),
as first proposed in Ref.~\cite{KaplanB}.
This idea is attractive from the point of view of $CP$ violation
and flavor-changing neutral currents (FCNCs) \cite{RC}.
Here, we will limit ourselves to a discussion of the top quark sector,
where the main concerns are to generate the experimentally measured value
of the top quark's mass naturally,%
\footnote{Without causing problems for $Z\to b\overline{b}$ decays
  \cite{ferretti,ACDP}.
}
together with a Higgs potential that triggers electroweak symmetry breaking.
It is generally acknowledged that the mass of the top quark sets it
apart from the other SM fermions as it is the only SM fermion with a mass
of the order of the electroweak symmetry breaking scale,
$v\sim 250$~GeV.   This suggests that the top quark may play an essential
role in generating electroweak symmetry breaking, whereas the origin
of the other SM fermion masses, and the strength of other symmetry breakings
such as $CP$ violation and FCNCs, might be very different.

The concrete hypercolor theory we will study in this article was proposed in
Ref.~\cite{ferretti}.  It was preceded by a general study
that highlighted what makes that theory particularly attractive \cite{FK}.%
\footnote{See also Ref.~\cite{Vecchi}.}
The hypercolor theory is a vector-like $SU(4)$ gauge theory
with fermions in two different \irreps\ (irreducible representations).
One of these \irreps, the six-dimensional two-index antisymmetric \irrep,
is real.  With 5 Majorana (or Weyl) fermions in this \irrep,
dynamical symmetry breaking in that sector of the theory
gives rise to an $SU(5)/SO(5)$ non-linear sigma model as its
low-energy effective theory.  As we will see in detail below,
the Higgs field lives in this non-linear sigma model.

Generally speaking, composite-Higgs models often rely on an
$SU(N_w)/SO(N_w)$ non-linear sigma model, which can arise from chiral symmetry
breaking in a theory containing $N_w$ Weyl (or Majorana) fermions in
a real \irrep.  An alternative coset structure is $SU(N_w)/Sp(N_w)$,
for which the $N_w$ Weyl fermions should be in a pseudoreal \irrep\ \cite{vac}.
For recent lattice work involving the pseudoreal fundamental \irrep\
of $SU(2)$ gauge theory we refer to Refs.~\cite{LPS,DCP}.  For a review
on beyond-the-SM lattice work, see Ref.~\cite{Kuti}.

The most familiar example of a real \irrep\ is the adjoint representation,
which occurs for example in supersymmetric theories \cite{susy}.
However, 5 Majorana fermions in the adjoint \irrep\ would most likely
push the theory from being confining to being conformal, even before
the introduction of any fermions in another \irrep.%
\footnote{See Ref.~\cite{Kuti} and references therein.}
Avoiding the adjoint \irrep,
the smallest instance of a real \irrep\ is the sextet of $SU(4)$.

Our goals are as follows.   First, a gauge theory
such as this $SU(4)$ hypercolor theory is amenable
to investigations using the methods of lattice gauge theory.
The effective theory below the hypercolor scale,
relevant for SM phenomenology, can be parametrized
in terms of low-energy couplings (LECs).
These LECs can be expressed in terms of correlation functions
in the hypercolor theory, which, in turn, allows for their computation
on the lattice.   While this is well understood
for the electroweak gauge sector, a similar careful derivation
of the LECs controlling the top sector has to our knowledge not been given
to date.  We derive the necessary correspondence
using spurion techniques.

Second,
once the connection between the effective theory and the hypercolor theory
has been established, we find that
it is possible to obtain semi-quantitative estimates of
the size of these LECs, using large-$N$ methods and factorization.
In particular, we show that the contribution of the top quark
to the Higgs effective potential indeed drives electroweak symmetry breaking
in a particular large-$N$ limit.

We expect that the techniques developed in this article can be easily extended
to similar hypercolor models.  In this sense, our choice of the model
of Ref.~\cite{ferretti} should be considered as a useful example.

This article is organized as follows.  Sec.~\ref{ferretti-model}
introduces and reviews the hypercolor theory \cite{ferretti},
including its field content, symmetries, and the effective non-linear fields
that will be needed for the low-energy effective theory.
In Sec.~\ref{weak-effpot} we briefly discuss the contribution of the
electroweak gauge bosons to the Higgs effective potential.
The main part of this article is Sec.~\ref{top},
where we discuss the top quark sector in detail.
We introduce the top quark spurions and the hyperbaryons in Sec.~\ref{baryons}.
We discuss the top Yukawa coupling in Sec.~\ref{top-mass},
and the top quark contribution to the Higgs effective potential
in Sec.~\ref{top-effpot}.   In Sec.~\ref{large-N} we define a large-$N$ limit
of the model, and show that for large enough $N$ the top-induced
Higgs potential will lead to electroweak symmetry breaking.
For simplicity, we assume a minimal explicit breaking
of the flavor group of the hypercolor theory by the couplings
to the SM.  In Sec.~\ref{pheno} we briefly comment on the more general
situation that arises if we relax this assumption.
In Sec.~\ref{lattice} we discuss the similarities between the
hypercolor theory and QCD, thus arguing that techniques developed to
study QCD on the lattice should be sufficient for hypercolor theories as well.
Section~\ref{conclusion} contains our conclusions.
A short appendix collects some of our conventions.

\section{\label{ferretti-model} Ferretti's model}
In Ref.~\cite{FK}, several requirements were put forward for a class of
composite Higgs models based on a hypercolor gauge theory as a UV completion.
We begin by listing these requirements.
The gauge group is assumed to be simple, and the dynamical symmetry
breaking pattern, $G\to H$, to be such that
\begin{eqnarray}
\label{dynsym}
  H &\supset& SU(3)_{\rm color}\times SU(2)_L\times SU(2)_R\times U(1)_X
\\
    &\supset& SU(3)_{\rm color}\times SU(2)_L\times U(1)_Y\ ,
\nonumber
\end{eqnarray}
with the SM gauge group in the last line.  The group
$SU(2)_R$ is the familiar custodial symmetry of the SM, and
the hypercharge is $Y=T^3_R+X$.  The SM Higgs doublet,
with quantum numbers $({\bf 1},{\bf 2},{\bf 2})_0$ under
$SU(3)_{\rm color}\times SU(2)_L\times SU(2)_R\times U(1)_X$, should be contained
in the NGB multiplet associated with the symmetry breaking $G\to H$.
In order to accommodate a partially composite top quark \cite{KaplanB},
\ie, for the top quark to acquire its mass through linear couplings
to hyperbaryons, there must exist hyperbaryons with quantum numbers
that match those of the SM quarks.
This includes a set of right-handed, spin-$1/2$ hyperbaryons
with quantum numbers $({\bf 3},{\bf 2})_{1/6}$ of the SM gauge group
$SU(3)_{\rm color}\times SU(2)_L\times U(1)_Y$, which serve
as partners of the SM quark doublet $q_L$;
and left-handed, spin-$1/2$ hyperbaryons
with the quantum numbers  $({\bf 3},{\bf 1})_{2/3}$, to serve as partners
of the SM quark singlet $t_R$.  Finally, the hypercolor theory
should be asymptotically free, and both the hypercolor gauge group and the
SM gauge group should be free of anomalies.

The hypercolor model with the smallest gauge group
that satisfies all these requirements is an $SU(4)$ gauge theory \cite{FK}.
The hyperfermion content consists of
five Majorana fermions $\c_i$, $i=1,\dots,5$, transforming in the
six-dimensional two-index antisymmetric \irrep\ of hypercolor,
which is a real representation; and three Dirac fermions $\j_a$, $a=1,2,3$,
in the fundamental representation.
The Majorana field $\c$ can be written in terms of a Weyl fermion $\U$ as
\begin{subequations}
\label{maj}
\begin{eqnarray}
  \c_{ABi} &=&
  \left(
  \begin{array}{c}
    \vspace{1ex}
    \U_{ABi} \\
    \half\e_{ABCD}\,\e\,(\bU^{CD}_i)^T
  \end{array}
  \right) \ ,
\label{maja}\\
  \bc^{AB}_i &=& \half \e^{ABCD}\c^T_{CDi} \, C \ = \ \rule{0ex}{3.5ex}
  \left(
  \begin{array}{cc}
    \! -\half\e^{ABCD}(\U_{CDi})^T \e  &  \bU^{AB}_i
  \end{array}
  \right) \ .
\label{majb}
\end{eqnarray}
\end{subequations}
We use capital letters for the $SU(4)$ hypercolor indices,
with lower indices for the fundamental \irrep, and upper indices
for the anti-fundamental \irrep.  Several lower or upper indices
will always be fully antisymmetrized.
A Dirac fermion $\j$ in the fundamental \irrep\ can be written in terms
of two right-handed Weyl fermions, $\J$ in the fundamental \irrep\
and $\tJ$ in the anti-fundamental, as
\begin{equation}
  \j_{Aa} = \left( \begin{array}{c}
    \J_{Aa} \\ \e \btJ^T_{Aa}
  \end{array} \right) \ , \qquad
  \bj_a^A = \left( -(\tJ_a^A)^T\e \ \ \bJ_a^A \right) \ .
\label{DW}
\end{equation}
We suppress spinor indices.  $C$ is the charge-conjugation matrix,
$\e=i\s_2$ is the two-dimensional $\e$-tensor acting on the Weyl
spinor index,
and the superscript $T$ denotes the transpose in spinor space.
With the lattice in mind, we work in euclidean space, choosing our Dirac
matrices to be hermitian and using the chiral representation, see App.~\ref{conv}.

The hypercolor theory possesses a flavor symmetry group
\begin{equation}
\label{flavor}
  G = SU(5)\times SU(3)\times SU(3)'\times U(1)_X\times U(1)'\ ,
\end{equation}
with quantum numbers $({\bf 5},{\bf 1},{\bf 1})_{(0,-1)}$ for $\U$;
$({\bf 1},{\bf \bar{3}},{\bf 1})_{(1/3,5/3)}$ for $\J$;
and $({\bf 1},{\bf 1},{\bf 3})_{(-1/3,5/3)}$ for $\tJ$.%
\footnote{Compare Table~1 of Ref.~\cite{ferretti}.}

We assume that dynamical symmetry breaking takes place,
generating a condensate $\langle\bc_i\c_j\rangle\propto\d_{ij}$
that breaks $SU(5)\to SO(5)$.
Consistent with the general considerations of Ref.~\cite{vac},
the Majorana bilinear $\bc_i\c_j$ is antisymmetric
on its spinor indices and symmetric on its hypercolor indices,
and so it is symmetric on its flavor indices.
In addition, there is a condensate $\langle\bj_a\j_b\rangle\propto\d_{ab}$
that breaks $SU(3)\times SU(3)'$ to its diagonal subgroup, which we
identify with $SU(3)_{\rm color}$.  Both condensates also break $U(1)'$.
The unbroken group is
\begin{equation}
\label{breaking}
  H = SO(5)\times SU(3)_{\rm color}\times U(1)_X\ .
\end{equation}
For heuristic arguments supporting this pattern of symmetry breaking,
see Refs.~\cite{vac,ferretti}.  Of course, whether this is the actual
symmetry breaking pattern is something that can be investigated
on the lattice.  Indeed the symmetry breaking pattern
of the Dirac fermions, with $SU(3)\times SU(3)'$ breaking to
the diagonal $SU(3)$ subgroup, is consistent with all known lattice results.
A first study of the real-\irrep\ symmetry breaking pattern,
in a similar theory except with four, instead of five, Majorana fermions,
has recently appeared in Ref.~\cite{taco}.

The effective theory at energy scales much below the hypercolor scale
$\L_{\rm HC}$ thus contains NGBs parametrizing the $U(1)'$ group manifold,
and the cosets $SU(3)\times SU(3)'/SU(3)_{\rm color}$ and $SU(5)/SO(5)$,
amounting to 1, 8 and 14 NGBs for each of these factors, respectively.
These NGBs are massless when all couplings of the hypercolor theory
to the SM are turned off.  A non-trivial effective potential
is induced both by the SM gauge bosons, as we briefly review
in Sec.~\ref{weak-effpot}, and by the coupling to the third-generation quarks.
The latter, which is the main subject of this paper, will be studied
in Sec.~\ref{top}.

The Higgs doublet is a subset of the NGB multiplet parametrizing the coset
$SU(5)/SO(5)$.  In more detail,
the 14 NGBs corresponding to the generators in this coset
are described by a non-linear field $\S\in SU(5)/SO(5)$
obtained by considering fluctuations
around the vacuum $\langle\S\rangle=\S_0= {\bf 1}$,
\begin{equation}
\label{Sigma}
  \S = u\,\S_0\,u^T = \mbox{exp}(i\P/f)\,\S_0\,\mbox{exp}(i\P/f)^T
  = \mbox{exp}(2i\P/f)\ ,
\end{equation}
with%
\footnote{Note that in Ref.~\cite{ferretti}, the notation $\S$ is used for the
  field $u$ of Eq.~(\ref{Sigma}).
}
\begin{equation}
\label{Sigmasymm}
  \S = \S^T\quad\Rightarrow\quad\P=\P^T\ .
\end{equation}
Under $g\in SU(5)$, $\S$ transforms as $\S\to g\S g^T$.

At the level of the algebra, $SU(2)_L\times SU(2)_R$ in Eq.~(\ref{dynsym}) is
equivalent to the $SO(4) \subset SO(5)$ associated with the first
four rows and columns.  The explicit form of the generators is given
in the appendix.  With this choice,
the field $\P$ can be written as
\begin{equation}
\label{Pi}
  \P = \Theta+\Theta^\dagger+\Phi_0+\Phi_++\Phi_+^\dagger+\eta\ ,
\end{equation}
with $\Theta$ containing the Higgs doublet $H=(H_+,H_0)^T$,
\begin{equation}
\label{H}
\Theta=\left(\begin{array}{ccccc}
0 & 0 & 0 & 0 & -iH_+/\sqrt{2} \\
0 & 0 & 0 & 0 & H_+/\sqrt{2} \\
0 & 0 & 0 & 0 & iH_0/\sqrt{2} \\
0 & 0 & 0 & 0 & H_0/\sqrt{2} \\
-iH_+/\sqrt{2} & H_+/\sqrt{2} & iH_0/\sqrt{2}
& H_0/\sqrt{2} & 0
\end{array}\right)\ .
\end{equation}
For the explicit parametrization of the rest of $\P$,
we refer to Ref.~\cite{ferretti}, as we will not need it here.
The Higgs doublet comprises four of the NGBs,
and the $SU(2)_L$ triplets $\phi_0$, $\phi_+$ and $\phi_-=(\phi_+)^\dagger$
comprise 9 more NGBs.%
\footnote{In the notation of Ref.~\cite{ferretti},
  $\phi_0=(\phi_0^-,\phi_0^0,\phi_0^+)$ with $\phi_0^0$ real and
  $\phi_0^+ = (\phi_0^-)^*$, while $\phi_+=(\phi_+^-,\phi_+^0,\phi_+^+)$,
  with all components complex.
}
Finally, $\P$ has a component $\eta$ proportional to
the generator $\mbox{diag}(1,1,1,1,-4)$, which is neutral with
respect to the entire SM model gauge group, and which completes the multiplet
of 14 NGBs.

While they play only a small role, we will need also
the non-linear fields associated with the other broken symmetries.
We account for the $SU(3)\times SU(3)'/SU(3)_{\rm color}$ coset
by a non-linear field $\O\in SU(3)$, transforming as $\O\to g\O h^\dagger$
for $g\in SU(3)$ and $h\in SU(3)'$.
A similar non-linear field arises in the familiar chiral lagrangian of
3-flavor QCD, but the reader should keep in mind the different
physical roles of the various $SU(3)$ groups in the case at hand.
Throughout most of this paper, we will assume that the only source
of \textit{explicit} breaking of $SU(3)\times SU(3)'$ to $SU(3)_{\rm color}$
arises from the coupling of the $SU(3)_{\rm color}$ currents to the SM gluons.
A wider range of possibilities than what is the main focus of this article
is allowed if we relax this assumption.
This is briefly discussed in Sec.~\ref{pheno}.
The 8 NGBs associated with the non-linear $\O$ field transform in
the adjoint \irrep\ of $SU(3)_{\rm color}$.
They are singlets under both $SU(2)_L$ and $U(1)_Y$.

Finally, to account for the spontaneous breaking of $U(1)'$
we introduce a non-linear field $\F\in U(1)$
with unit charge under $U(1)'$.
The associated NGB, $\eta'$, is neutral under the SM gauge interactions.
Using a $\sim$ sign to indicate
identical transformation properties under the entire flavor group $G$,
we thus have
\begin{subequations}
\label{transG}
\begin{eqnarray}
\label{transGa}
  \F^{-2}\, \S_{ij} &\sim& \bc_i P_R \c_j
  \ \sim \ \e^{ABCD} (\U_{CDi})^T \,\e\, \U_{ABj} \ ,
\\
\label{transGb}
  \F^{-10/3}\, \O_{ab} &\sim& \bj_a P_L \j_b
  \ \sim \ \bJ_a^A \,\e\, (\btJ_{Ab})^T \ .
\end{eqnarray}
\end{subequations}

\section{\label{weak-effpot} Higgs effective potential from electro-weak gauge bosons}
In this section, we briefly review the contribution
from the SM gauge bosons to the effective potential for the NGBs,
starting with the effective potential for
the $SU(5)/SO(5)$ non-linear field $\S$ generated by the electroweak
gauge bosons.  This part of the effective potential takes the form
\begin{equation}
\label{VeffEW}
  V_{\rm eff}^{\rm EW}(\S) = C_{LR}\sum_Q\tr\left(Q\S Q^*\S^*\right)\ ,
\end{equation}
if we work to leading (\ie, quadratic) order in the SM gauge couplings.
The sum over $Q$ runs over the $SU(2)_L$ generators
$gT^a_L$ with $T^a_L$ given in Eq.~(\ref{embed}),
and the hypercharge generator $g'Y=g'\left(T^3_R+X\right)$,
with $X=0$ for the $\P$ field.  Here
\begin{equation}
\label{CLR}
  C_{LR} = \frac{3}{(4\p)^2}\int_0^\infty dq^2\,q^2\,\P_{LR}(q^2)\ ,
\end{equation}
and
\begin{equation}
\label{PiLR}
  \left(q^2\d_{\m\n}-q_\m q_\n\right)\P_{LR}(q^2)
  = \int d^4x\,e^{iqx}\,
  \tr \svev{\g_\m P_R[\c(x)\bc(0)]\g_\n P_L[\c(0)\bc(x)]} \ ,
\end{equation}
where $[\c(x)\bc(y)]$ is the Majorana fermion propagator for the field $\c$
of Eq.~(\ref{maj}), and $\langle\dots\rangle$ indicates the expectation value with
respect to the hypercolor gauge fields.  This type of effective potential
goes back to the well-known formula for the mass difference
between the charged and neutral pions in QCD.
For further explanations and a derivation of this result in the present
context we refer to the review article Ref.~\cite{RC}
and to the appendix of Ref.~\cite{vacalignW}.%
\footnote{The factor of 3 in Eq.~(\ref{CLR}) comes from tracing over the transversal
projector.  This factor is erroneously missing in the published versions
of Eqs.~(A11a) and~(A11b) of Ref.~\cite{vacalignW}, and Eq.~(A5) of
Ref.~\cite{vacalignS}.  Also, in the published versions,
the term $gW_\mu Q_a J_{\mu A}$ in Eq.~(A2) of Ref.~\cite{vacalignW}
should be multiplied by $i$, and similarly for the term
$gW_\mu (Q_a^L J_{\mu a}^L + Q_a^R J_{\mu a}^R)$ in Eq.~(A2) of Ref.~\cite{vacalignS}.
All these factors have been corrected in the current archive versions.
}

The proof of Ref.~\cite{EW} that $C_{LR}>0$ applies also in this case.
Using the explicit form~(\ref{embed}),
the minimum of $V_{\rm eff}^{\rm EW}(\S)$ is equal to
$-C_{LR}(3g^2+g'^2)$, which is attained at $\S={\bf 1}$.  This
part of the effective potential does not rotate the vacuum of the hypercolor
theory, exhibiting the phenomenon of vacuum alignment \cite{vac}.

Expanding $\S$ to quadratic order in $\P$ inside $V_{\rm eff}^{\rm EW}$,
we find the mass terms
\begin{equation}
\label{massNGBsEW}
  V_{\rm eff}^{{\rm EW}(2)} = \frac{C_{LR}}{f^2}
  \left\{(3g^2+g'^2)\left(2H^\dagger H+\frac{16}{3}\,\phi_+^\dagger
         \phi_+\right)+8g^2\phi_0^\dagger\phi_0\right\}\ .
\end{equation}
The explicit symmetry breaking by
the electroweak gauge bosons produces a positive mass-squared for all components
of the NGB multiplet $\P$ except $\eta$.

To summarize, when we couple the hypercolor theory to the SM gauge bosons,
an effective potential for the non-linear field $\S$ is generated.
At lowest order, it is proportional to the squares of the electroweak couplings
$g_{\rm EW}=g$ for $SU(2)_L$, or $g_{\rm EW}=g'$ for $U(1)_Y$,
where it is understood that all SM couplings are evaluated
at the hypercolor scale $\L_{\rm HC}$.
Because of vacuum alignment, the expectation value $\S_0$ will remain
equal to one, but the Higgs doublet and the three $SU(2)_L$ triplets
will acquire a mass proportional to $g_{\rm EW}f$, while the
singlet $\eta$ will remain massless. To avoid confusion,
the contribution to the effective potential from the top quark has not yet
been included,
and will be discussed in the next section.

Similarly, when we turn on the QCD interactions,
an effective potential for the $\O$ non-linear field is generated,
\begin{equation}
\label{VeffQCD}
  V_{\rm eff}^{\rm QCD}(\O) = -C_{LR}^{\rm QCD}\sum_Q\tr\left(Q\O Q\O^\dagger\right)\ ,
\end{equation}
where now $Q$ runs over the 8
generators $g_{\rm s}\l_a$ of $SU(3)_{\rm color}$,
and $g_{\rm s}$ is the QCD coupling (at the hypercolor scale).
In the underlying hypercolor theory, $C_{LR}^{\rm QCD}$ has
a representation analogous to Eq.~(\ref{CLR}),
except that the Majorana-fermion propagator in Eq.~(\ref{PiLR})
is replaced by the Dirac-fermion propagator $[\j(x)\bj(y)]$.
Once again there is vacuum alignment, nailing down the vacuum at
$\svev{\O_{ab}}=\d_{ab}$, and giving the octet of NGBs a mass
of order $g_{\rm s}f$.  Thus,
as long as $f$ is much larger than the electroweak scale,
both the $SU(2)_L$ triplet NGBs and the color-octet NGBs are much heavier
than the electroweak gauge bosons or the top quark.

As already noted,
the NGB $\eta'$ of the spontaneously broken $U(1)'$ is inert
under all the SM gauge interactions.  Moreover, the coupling of the
hypercolor sector to the SM considered in the next section does not break
$U(1)'$ explicitly.  Therefore, no effective potential will be generated for
the associated non-linear field $\F$, and $\eta'$ will remain exactly
massless.

\section{\label{top} The top quark sector}
We now proceed to the main part of this article,
which is the study of the dynamics arising from the coupling
of the top quark to the hypercolor theory.
There are two aspects of interest: the contribution of the top quark to the
effective potential for the non-linear field $\S$ containing
the Higgs field, analogous to the
contribution from the weak gauge bosons in Eq.~(\ref{VeffEW}); and the mass
of the top quark itself.

We will begin with the coupling of the top quark to the hypercolor theory at
the ``microscopic'' level, which involves only the elementary fields:
the hypercolor gauge fields, and the hyperfermions of Eqs.~(\ref{maj}) and~(\ref{DW}).
We introduce fermionic spurions which transform in complete
representations of the flavor group $G$ of Eq.~(\ref{flavor}), and which contain the
$SU(2)_L$ doublet $q_L$ of the left-handed top and bottom quarks
$t_L$ and $b_L$, as well as the right-handed top quark $t_R$.
Following the mechanism proposed in Ref.~\cite{KaplanB},
the spurions will be coupled linearly to suitable hyperbaryon fields,
which are three-fermion operators in the hypercolor theory.
We demand that the spurion--hyperbaryon interactions
are invariant under $G$, because the microscopic theory does not
know about the dynamical breaking $G\to H$.

The spurion--hyperbaryon interaction terms are
four-fermion operators, and are assumed to arise from some extended
hypercolor (EHC) sector with a dynamical scale $\L_{\rm EHC}\gg
\L_{\rm HC}$, the origin of which we will not specify.
We will return to this point in the conclusion section.
The hyperbaryon operators and the four-fermion couplings
are constructed in Sec.~\ref{baryons}.

We then turn to the effective low-energy theory.
We demand that also the effective theory is invariant under $G$,
but it can now depend on the effective fields:
the $SU(5)/SO(5)$ coset field $\S$, which plays
a central role since it contains the Higgs field,
as well as the $SU(3)\times SU(3)'/SU(3)_{\rm color}$ field $\O$
and the $U(1)$-valued field $\F$.
Note that we do not allow the effective theory to contain
any effective fields for hyperbaryons.  This strategy generalizes
the standard construction of the chiral lagrangian for QCD.%
\footnote{See for instance Ref.~\cite{MGchpt}.}

We proceed in two steps.  First, in Sec.~\ref{top-mass}, we consider the
coupling of the SM quarks $q_L$ and $t_R$ to the effective non-linear fields,
integrating out all other states in the hypercolor theory.
This will lead to an expression for the top Yukawa coupling in terms of
a hyperbaryon two-point function in the hypercolor theory.

Next, in Sec.~\ref{top-effpot}, we consider the contribution
to the effective potential obtained by integrating also
over the third-generation quarks to leading order in the top Yukawa coupling.
This involves restricting the spurions to their SM values,
in which all components except those corresponding to
$q_L=(t_L,b_L)$ and to $t_R$ are set equal to zero, and
integrating over $q_L$ and $t_R$.
Like the coupling to the SM gauge bosons (Sec.~\ref{weak-effpot}),
this breaks explicitly the flavor group $G$.
However, in the approximation in which we work, only $SU(5)$ is broken explicitly,
whereas all other factors in Eq.~(\ref{flavor}) are not.
As a result, no effective potential is generated for $\O$ or $\F$.

\begin{table}[t]
\vspace*{3ex}
\begin{center}
\begin{tabular}{ c | clccc | c } \hline
  & $SU(5)$ & $SU(3)\times SU(3)'$ & $SU(3)_c$ & $U(1)_X$ & $U(1)'$ &
\\ \hline\hline
  $\U (\J\J)$      & \five    & $(\threebar,\one)\times(\threebar,\one)\to(\three,\one)$
  & \three    &  2/3 &   7/3 & $B_R$ \\
  $\U (\btJ\btJ)$  & \five    & $(\one,\threebar)\times(\one,\threebar)\to(\one,\three)$
  & \three    &  2/3 & -13/3 & $B'_R$ \\
  $\bU (\bJ\bJ)$   & \fivebar & $(\three,\one)\times(\three,\one)\to(\threebar,\one)$
  & \threebar & -2/3 &  -7/3 & $\bB_R$ \\
  $\bU (\tJ\tJ)$   & \fivebar & $(\one,\three)\times(\one,\three)\to(\one,\threebar)$
  & \threebar & -2/3 &  13/3 &  $\bB'_R$ \\
\hline\hline
  $\bU (\J\J)$     & \fivebar & $(\threebar,\one)\times(\threebar,\one)\to(\three,\one)$
  & \three    &  2/3 &  13/3 & $B_L$ \\
  $\bU (\btJ\btJ)$ & \fivebar & $(\one,\threebar)\times(\one,\threebar)\to(\one,\three)$
  & \three    &  2/3 &  -7/3 & $B'_L$ \\
  $\U (\bJ\bJ)$    & \five    & $(\three,\one)\times(\three,\one)\to(\threebar,\one)$
  & \threebar & -2/3 & -13/3 & $\bB_L$ \\
  $\U (\tJ\tJ)$    & \five    & $(\one,\three)\times(\one,\three)\to(\one,\threebar)$
  & \threebar & -2/3 &   7/3 & $\bB'_L$ \\
\hline\hline
\end{tabular}
\end{center}
\vspace*{-3ex}
\begin{quotation}
\floatcaption{tabHC}{%
Local hyperbaryon operators.   The leftmost column gives the Weyl-fermion
content, and the rightmost column the notation used for the operator.
The remaining columns list the quantum numbers.
}
\end{quotation}
\vspace*{-4.5ex}
\end{table}

The explicit breaking of $SU(5)$ generates an effective potential for $\S$.
This new contribution is parametrized by one new LEC,
$C_{\rm top}$, analogous to $C_{LR}$ in Eq.~(\ref{VeffEW}). We will show
that $C_{\rm top}$ can be expressed as an integral over a hyperbaryon
four-point function convoluted with two free, massless fermion propagators.

Up to this point, our analysis is from first principles.
In Sec.~\ref{large-N} we turn to physical but non-rigorous considerations.
We show that
a large-$N$ limit exists (where $N=4$ for the hypercolor group $SU(4)$ we
consider here) in which the hyperbaryon four-point function factorizes,
leading to a simple result for $C_{\rm top}$, and ultimately to a
non-trivial expectation value for the Higgs field.

Finally, in Sec.~\ref{pheno} we comment on the phenomenological consequences
of our analysis.  This includes a brief discussion of the more general
situation where the explicit breaking of $SU(3)\times SU(3)'$
to its diagonal subgroup $SU(3)_{\rm color}$ is allowed to come from
other sources than the QCD gluons.

\subsection{\label{baryons} Top quark spurions and hyperbaryons}
We begin with introducing the top-quark spurions,
a left-handed spurion $T_L$, which we choose in the {\bf 5} \irrep\ of $SU(5)$,
and a right-handed spurion $T_R$, which we choose
in the ${\bf \bar{5}}$ \irrep.  Both \irreps\ reduce to the
{\bf 5} of $SO(5)$.  This choice ensures that
terms like $\bT_L T_R$ in the effective potential are disallowed by
$SU(5)$, but allowed by $SO(5)$.%
\footnote{We may switch  {\bf 5} with ${\bf \bar{5}}$, but the key point
  is that the two spurions are chosen to be in different $SU(5)$ \irreps.
}
Both $T_L$ and $T_R$ will have $U(1)_X$ charge $2/3$,
as this will yield the correct hypercharges for $q_L$ and $t_R$.
The SM values for these spurions are
\begin{equation}
\label{fspurions}
  T_L = \hT_L \equiv \frac{1}{\sqrt{2}}\left(\begin{array}{c}
    ib_L\\b_L\\i t_L\\-t_L\\0
  \end{array}\right)\ ,\qquad
  T_R = \hT_R \equiv \left(\begin{array}{c}
    0\\0\\0\\0\\i t_R
  \end{array}\right)\ .
\end{equation}
Using Eq.~(\ref{embed}), it is straightforward to verify that the pair $(t_L,b_L)$
transforms as an $SU(2)_L$ doublet, and has hypercharge $Y=1/6$.
The $SU(2)_L$ singlet $t_R$ has hypercharge $Y=2/3$.
The quantum numbers of the spurions under the remaining flavor symmetries
will be discussed shortly.

The spurions couple to the hypercolor theory through
the $G$-invariant lagrangian
\begin{equation}
\cl_{\rm EHC} =  \l_1\bT_L B_R +\l_1^*\bB_R T_L
  + \l_2 \bT_R B_L + \l_2^*\bB_L T_R\ ,
\label{lagSU5}
\end{equation}
where $B_{L,R}$ are hyperbaryon fields with appropriate
quantum numbers.   Setting the spurions $T_{L,R}$
equal to their SM values~(\ref{fspurions}) then tells us how the SM quarks
$t_L$, $b_L$ and $t_R$ couple to the hypercolor theory.
Since we will use three-hyperfermion local interpolating fields
for the hyperbaryons, the four-fermion interactions in $\cl_{\rm EHC}$ have
engineering dimension six.  $\cl_{\rm EHC}$ originates from some other
sector with scale $\L_{\rm EHC}\gg\L_{\rm HC}$, with effective couplings
$\l_{1,2}\sim O(\L_{\rm EHC}^{-2})$ just below that scale.

Limiting ourselves to local hyperbaryon fields,
all operators that can be used in the construction of a $G$-invariant
$\cl_{\rm EHC}$ are listed in Table~\ref{tabHC}.
The schematic structure in terms of Weyl fields
is indicated in the first column of the table, followed by the quantum numbers
under the flavor group $G$.  The column labeled as $SU(3)_c$
gives the $SU(3)_{\rm color}$ \irrep.  The spinor index of the hyperbaryon
field is always carried by the Majorana fermion $\c$.
Using Eqs.~(\ref{maj}) and~(\ref{DW}), explicit expressions for the
unprimed operators in Table~\ref{tabHC} are
\begin{subequations}
\label{bar4}
\begin{eqnarray}
  B_{Ria} &=& -\half \e^{ABCD} \e_{abc}\, P_R\, \c_{ABi}\,
                  \left( \j_{Cb}^T \,C P_R\, \j_{Dc} \right)
\label{bar4a}\\
  &=& \half \e^{ABCD} \e_{abc}\, \U_{ABi}
      \left(\J^T_{Cb}\,\e\,\J_{Dc}\right) \ ,
\nonumber\\
  \bB_{Ria} &=& \half \e_{ABCD} \e_{abc}\, \bc^{AB}_i P_L\,
                  \left( \bj^C_b C P_L \Big(\bj^D_c\Big)^T \right)
\label{bar4b}\\
  &=& \half \e_{ABCD} \e_{abc}\, \bU^{AB}_i
      \left(\bJ^C_b\,\e\,\Big(\bJ^D_c\Big)^T\right) \ ,
\nonumber\\
  B_{Lia} &=& -\half \e^{ABCD} \e_{abc}\, P_L \c_{ABi}\,
                  \left( \j_{Cb}^T \,C P_R\, \j_{Dc} \right)
\label{bar4c}\\
  &=& \e_{abc}\,\e\,\Big(\bU^{AB}_i\Big)^T
      \left(\J^T_{Ab}\,\e\,\J_{Bc}\right) \ ,
\nonumber\\
  \bB_{Lia} &=& \half \e_{ABCD} \e_{abc}\, \bc^{AB}_i P_R\,
                  \left( \bj^C_b C P_L \Big(\bj^D_c\Big)^T \right)
\label{bar4d}\\
  &=& \e_{abc}\,\U^T_{ABi}\,\e
      \left(\bJ^A_b\,\e\,\Big(\bJ^B_c\Big)^T\right) \ .
\end{eqnarray}
\end{subequations}
The primed operators in Table~\ref{tabHC} are obtained from Eq.~(\ref{bar4})
by interchanging $P_R\leftrightarrow P_L$ inside the $\j\j$
and $\bj\bj$ bilinears.

When the spurions $T_{L,R}$ are restricted to their SM values,
the phases of $\l_{1,2}$ in Eq.~(\ref{lagSU5})
can be removed by (non-anomalous) $SU(2)_L$ and
$SU(2)_R$ transformations on the spurion fields,
implying that, from now on, we may take $\l_{1,2}$ to be real and positive.
This allows us to require that the lagrangian~(\ref{lagSU5}) be $CP$ invariant.
The $CP$ transformation acts as
\begin{equation}
  \j \to \g_2 \bj^T \ , \qquad \bj \to \j^T \g_2 \ ,
\label{CP}
\end{equation}
for both Dirac and Majorana fermions (see App.~\ref{conv} for our Dirac matrices
conventions).  The sign choices we have made in Eq.~(\ref{bar4})
imply that a $CP$ transformation applied
to the elementary fields $\c$, $\j$ and $\bj$ induces a $CP$ transformation
of the same form
on the hyperbaryon fields as well, thereby ensuring the $CP$ invariance
of $\cl_{\rm EHC}$.

The unprimed fields in Table~\ref{tabHC} transform non-trivially
under $SU(3)$ and are singlets under $SU(3)'$, whereas for the primed fields
the opposite is true.  Choosing either the primed or the unprimed version
for each hyperbaryon field gives rise to a total of four different
possibilities for $\cl_{\rm EHC}$.
The quantum numbers of the spurions $T_L$ and $T_R$
are chosen accordingly, so as to ensure the $G$ invariance of $\cl_{\rm EHC}$.
The SM quark fields $q_L$ and $t_R$ are endowed with same
$SU(3)\times SU(3)'\times U(1)_X\times U(1)'$ quantum numbers
as their parent spurion.  This construction is consistent with the SM,
since the resulting quantum numbers under $SU(3)_{\rm color}$
will always be the same.  In addition, it follows that
the entire theory, including the hypercolor sector, the SM lagrangian,
and their coupling via $\cl_{\rm EHC}$, is invariant under both $SU(3)$
and $SU(3)'$, provided that the QCD interactions can be neglected.
This is indeed the case in this section, because we calculate the Higgs
effective potential to second order in all SM couplings,
and since the result will be quadratic in the top Yukawa coupling,
any corrections that involve an additional dependence
on the QCD coupling $g_{\rm s}$ are neglected.

Note that we cannot generalize Eq.~(\ref{lagSU5}) to include,
simultaneously, terms that couple a given spurion to both of
the unprimed and primed hyperbaryons, as this will not allow for any
consistent assignment of $SU(3)\times SU(3)'$ quantum numbers.
For example, $\bT_L$ can couple to either $B_R$ or $B'_R$, but not to both.
It is because of this fact that $\cl_{\rm EHC}$ depends on only two
coupling constants $\l_{1,2}$.  This will lead to considerable simplification
in our analysis.  In Sec.~\ref{pheno} we briefly comment on the
more general case, where $\cl_{\rm EHC}$ is restricted only by
$SU(3)_{\rm color}$.

\subsection{\label{top-mass} The top quark Yukawa coupling}
Our next task is to construct the electroweak effective field theory.
As a first step, we integrate only over the gauge fields and fermions
of the hypercolor theory, and obtain an effective theory that depends on the
spurions $T_L$ and $T_R$, and on the non-linear fields,
including in particular the $SU(5)/SO(5)$ field $\S$.
We assume that the electroweak scale $m_W\sim m_t$ is much smaller
than the hypercolor scale $f\sim M\sim\L_{\rm HC}$, where $M$ is
of order the mass of the hyperbaryons which are assumed
to couple to the top quark in Eq.~(\ref{lagSU5}).   This provides us with
a power counting, and, in particular, the effective theory can be
organized according to a derivative expansion.   We will be concerned
with the lowest non-trivial order in this expansion.

Demanding full $G$ invariance, the leading order spurion potential is
\begin{equation}
\label{topmasseff}
  V_{\rm top} = \m_L\F^2\,\bT_R\S^*T_L + \m_R\F^{-2}\,\bT_L\S T_R\ .
\end{equation}
Terms like $\bT_L T_L$ vanish because of chiral projectors,
while terms like $\bT_L\bT_R^T$ are not allowed by $U(1)_X$ symmetry.
Bilinear terms independent of $\S$ are possible,
but thanks to $SU(5)$ invariance, they
have an $LL$ or $RR$ structure, and need an insertion of $\g_\m$.
Therefore, they contain at least one derivative, and their role is to
renormalize the kinetic terms for the top and bottom quarks,
which are present when these SM fields are made dynamical.
It can be checked that the correction is of order $y$,
where $y$ is the top quark Yukawa coupling introduced in Eq.~(\ref{y}) below.

$V_{\rm top}$ depends on two effective fields, $\S$ and $\F$.
The role of $\F$ is to reinstate $U(1)'$ invariance ({\it cf.} Eq.~(\ref{transG})).
When we choose both hyperbaryons in Eq.~(\ref{lagSU5}) to be unprimed ones,
the hyperbaryons and the spurions
transform non-trivially only under $SU(3)$, and are singlets of $SU(3)'$.
Therefore, $V_{\rm top}$ is invariant under $SU(3)\times SU(3)'$ as it stands,
without having to introduce any dependence on the effective field $\O$.

If we set $\S=\F=1$, and substitute the SM values
$\hT_L$ and $\hT_R$ defined in Eq.~(\ref{fspurions}) for $T_L$ and $T_R$,
we find that $V_{\rm top}$ vanishes.
$\S$ will need to develop a non-trivial expectation value for the SM
top quark to acquire a non-zero mass.
This will be discussed in Sec.~\ref{top-effpot} below.

In order to find the LECs $\m_{L,R}$ in Eq.~(\ref{topmasseff}),
we consider the second derivatives
\begin{equation}
\label{LRder}
  \frac{\partial^2}{\partial T_L(y)\partial\bT_R(x)}\log{Z}\ ,\qquad
  \frac{\partial^2}{\partial T_R(y)\partial\bT_L(x)}\log{Z}\ ,
\end{equation}
where $Z$ is the partition function of either the effective or
the microscopic theory.  Requiring the effective theory to match
the microscopic theory (and noting that the fermionic spurions are Grassmann)
yields the relations
\begin{eqnarray}
\label{mus}
  -\m_L P_L \langle\F^2\S^*\rangle\d(x-y) + \cdots
  &=&\l_1\l_2 P_L\langle B_L(x)\bB_R(y)\rangle P_L\ ,
\\
  -\m_R P_R \langle\F^{-2}\S\rangle\d(x-y) + \cdots
  &=&\l_1\l_2 P_R\langle B_R(x)\bB_L(y)\rangle P_R\ .
\nonumber
\end{eqnarray}
The ellipses on the left-hand side indicate that
the leading-order low-energy theory given by $V_{\rm top}$
reproduces the correlation functions on the right-hand side only to leading
order in a derivative expansion.

By assumption, symmetry breaking in the hypercolor theory yields
$\svev{\bc_i\c_j}\propto\svev{\S_{ij}}=\d_{ij}$,
up to a symmetry transformation.  (When the SM fields become dynamical
we may in general have $\svev{\S_{ij}}\ne\d_{ij}$, but these
corrections are of higher order in the SM gauge and Yukawa couplings.)
Setting $\svev{\S_{ij}}=\d_{ij}$ and $\svev{\F}=1$
in Eq.~(\ref{mus}) provides us with expressions for the parameters $\m_{L,R}$.
Assembling the chiral baryon fields together as
\begin{equation}
\label{deftB}
  B = B_R+B_L\ , \qquad \bB = \bB_R + \bB_L \ ,
\end{equation}
where $B$ is a Dirac field with quantum numbers $({\bf 5},{\bf 3})$
under the unbroken $SO(5)\times SU(3)_{\rm color}$, and writing
\begin{eqnarray}
\label{FT}
  \d(x-y) &=& \int\frac{d^4p}{(2\p)^4}\,e^{ip(x-y)}\ ,
\\
  \svev{B(x)\bB(y)}
  &=& \int\frac{d^4p}{(2\p)^4}\,e^{ip(x-y)}S_{B}(p)\ ,
\nonumber
\end{eqnarray}
we find,
to leading order in the momentum expansion of the effective theory,
\begin{eqnarray}
\label{lecs}
  \m_L P_L &=& -\l_1\l_2 P_LS_{B}(0)P_L\ ,
\\
  \m_R P_R &=& -\l_1\l_2 P_RS_{B}(0)P_R\ .
\nonumber
\end{eqnarray}
Apart from the chiral projectors, these two expressions must be equal, because
any hyperbaryon fields occurring in Eq.~(\ref{lagSU5}) can only have
a Dirac mass; Majorana masses (such as $B_R^T \e B_R$ or $B_L^T \e B_L$)
are forbidden by $U(1)_X$ symmetry.
Therefore,
\begin{equation}
\label{mu}
  \m=\m_L=\m_R=-\l_1\l_2 S_{B}(0)\ ,
\end{equation}
where we have used that at zero momentum $S_{B}(0)$ is proportional to
the unit matrix in spinor space.

We may now introduce the top quark Yukawa coupling $y$ by writing
\begin{equation}
  \m=yf/2 \ ,
\label{y}
\end{equation}
with $f$ the decay constant of the hypercolor theory.
If the Higgs field will now develop a non-zero expectation value,
\begin{equation}
\label{Hh}
  \langle H_0\rangle = \langle H_0^\dagger\rangle = h/\sqrt{2} \ ,
\end{equation}
this will induce a top quark mass
\begin{equation}
\label{topmass}
  m_t = \frac{1}{2\sqrt{2}}\,yf\sin{(2h/f)}\approx \frac{1}{\sqrt{2}}\,yh\ ,
\end{equation}
where we have used Eqs.~(\ref{Sigma}),~(\ref{H}),~(\ref{fspurions})
and~(\ref{topmasseff}), and the approximate equality holds for $h/f\ll 1$.

We may introduce an effective hyperbaryon field $\tB$ which is
canonically normalized by writing
\begin{equation}
\label{B}
  B = f^3\sqrt{Z_B} \tB\ .
\end{equation}
We define $M$ as the zero-momentum mass of the canonically normalized
$\bB$ field, corresponding to a term $M\btB \tB$.
In other words, $S_{B}(0) = f^6 Z_B/M$. This gives
\begin{equation}
\label{yukawa}
  y =   -2\l_1\l_2\,Z_B\,f^5/M\ .
\end{equation}
We comment that the field $B$ does not necessarily correspond to any
baryon mass eigenstate of the hypercolor theory.  Still,
generically we might expect it to couple to the lightest
hyperbaryon with quantum numbers that match those of the SM quarks,
in which case $M$ will be a quantity of the order of this smallest
hyperbaryon mass.

We conclude this subsection with a technical comment.
If in Eq.~(\ref{lagSU5}) we choose one unprimed and one primed hyperbaryon field,
this implies that one of the spurions transforms non-trivially
under $SU(3)$ while the other under $SU(3)'$.
In this case, $V_{\rm top}$ will depend on $\O$.
For definiteness, replacing $B_L$ in Eq.~(\ref{lagSU5}) by $B'_L$
implies that now $T_R$ transforms non-trivially under $SU(3)'$,
and Eq.~(\ref{topmasseff}) gets replaced by
\begin{equation}
\label{veffprimea}
  V_{\rm top} = \m_L\F^{-14/3}\,\bT_R \S^* \O^\dagger T_L
  + \m_R\F^{14/3}\,\bT_L\S \O T_R\ .
\end{equation}
This hardly changes our analysis, because,
in order to obtain expressions for the parameters $\m_{L,R}$
in Eq.~(\ref{topmasseff}) we are setting all non-linear fields
equal to the identity anyway.

In the next subsection, we will work out
the effective potential after integrating over the third-generation quarks.
In this calculation, any dependence on both $\O$ and $\F$ will drop out
regardless of our choice of hyperbaryon fields in Eq.~(\ref{lagSU5}), as it must be
because both $SU(3)\times SU(3)'$ and $U(1)'$ are not explicitly broken
in the top sector to the order we work, and therefore no effective potential
can be generated for those non-linear fields.
In particular, when $V_{\rm top}$ depends on $\O$ as in Eq.~(\ref{veffprimea}),
then expression~(\ref{vefftermsc}) below,
which is the only term that will contribute to the effective potential,
gets multiplied by $\tr(\O\O^\dagger)=\tr\,{\bf 1}$, showing that indeed
the $\O$ dependence cancels out.

\subsection{\label{top-effpot} Higgs effective potential induced by the top quark}
We now integrate over the SM top quark in order to obtain the associated
contribution $V_{\rm eff}^{\rm top}(\S)$ to the effective potential.
Adding this to Eq.~(\ref{VeffEW}) gives the complete effective potential
for $\S$ to second order in the SM gauge and Yukawa couplings.
We will disregard all the other SM fermions, including the bottom quark,
on the grounds that their Yukawa couplings are much smaller,
and so their contribution to the
effective potential will be much smaller as well.

We begin by splitting the spurions $T_{L,R}$ as follows%
\footnote{Whether or not $t_L$ in Eq.~(\ref{fsrewrite}) coincides with
  the component with the same name of $\hT_L$ in Eq.~(\ref{fspurions}) depends
  on the value we choose for $v_L$, as we will see below.
}
\begin{equation}
\label{fsrewrite}
  T_L(x) = t_L(x) v_L\ ,\qquad T_R(x) = t_R(x) v_R\ .
\end{equation}
The new global spurions $v_{L,R}$ carry the $SU(5)$ quantum numbers, which
contains the SM symmetry $SU(2)_L$.  We also assign $U(1)_X$ to these spurions,
because the hypercharge $Y$ is the sum of the charge $X$ and the third component
of $SU(2)_R$, with the latter being a subgroup of $SU(5)$ as well.
The (Grassmann) fields $t_{L,R}$ carry the spin index.
They also inherit the $SU(3)$ and $SU(3)'$ quantum numbers from $T_{L,R}$.
We promote $t_{L,R}$ to dynamical fields by adding tree-level kinetic terms
$\bt_L {\sl\partial} t_L + \bt_R {\sl\partial} t_R$.

The effective potential $V_{\rm eff}^{\rm top}$ at order $y$
is obtained by substituting Eq.~(\ref{fsrewrite}) into $V_{\rm top}$
of Eq.~(\ref{topmasseff}), and integrating over the top quark, leading to
a contribution with the form of $\F^{-2} \bv_L \S v_R + \mbox{h.c.}$.
This contribution vanishes, however,
because the only non-zero tree-level top propagators
are $\svev{t_L \bt_L}$ and $\svev{t_R \bt_R}$.

The leading contribution to $V_{\rm eff}^{\rm top}$ is of order $y^2$.
It involves four global spurions.
Momentarily suppressing any dependence on the $\O$ and $\F$ fields,
the possible terms that depend on $\S$ are
\begin{subequations}
\label{veffterms}
\begin{eqnarray}
\label{vefftermsa}
  (\bv_L\S v_R)^2+\mbox{h.c.}\ ,
\\
\label{vefftermsb}
\quad (\bv_R\S^* v_L)^2+\mbox{h.c.}\ ,
\\
\label{vefftermsc}
  (\bv_L\S v_R)(\bv_R\S^* v_L)\ .
\end{eqnarray}
\end{subequations}
The tree-level top propagators allow for the generation of the last term only.
If the effective potential~(\ref{vefftermsc}) arises as the product
of the two interactions in Eq.~(\ref{topmasseff}),
the $\F$ dependence evidently cancels out.  Moreover, as we have explained in
the previous section, regardless of the choice of hyperbaryon fields
we make in Eq.~(\ref{lagSU5}), $V_{\rm eff}^{\rm top}$ will be
independent of $\O$ and $\F$, because, to the order we are working,
$SU(3)$, $SU(3)'$ and $U(1)'$ are not broken explicitly.

Promoting the fields $t_L$ and $t_R$ in Eq.~(\ref{fspurions})
to be dynamical amounts to setting
\begin{equation}
\label{vLvR}
  v_L = \hv_L \equiv \frac{1}{\sqrt{2}}\left(\begin{array}{c}
    0\\0\\i \\-1\\0
  \end{array}\right)\ ,\qquad
  v_R = \hv_R \equiv \left(\begin{array}{c}
    0\\0\\0\\0\\i
  \end{array}\right)\ ,
\end{equation}
with $\hbv_{L,R} = \hv_{L,R}^\dagger$.
The resulting contribution to the effective potential is
\begin{equation}
\label{vtopeff}
  y^2 C_{\rm top}\,(\hbv_L\S \hv_R)(\hbv_R\S^* \bv_L)
  = \frac{y^2}{2}\,C_{\rm top}
  \left(\S_{35}-i\S_{45}\right) \left(\S_{35}^*+i\S_{45}^*\right)\ .
\end{equation}
$C_{\rm top}$ is a new LEC.  We have factored out the square of
the Yukawa coupling $y$ to make explicit the order at which we work.

However, we are not done yet.  In order to arrive at this result
we have used Eq.~(\ref{vLvR}) for the global spurions, which projects onto
a particular component of the $SU(2)_L$ doublet $q_L$, the one denoted $t_L$
in Eq.~(\ref{fspurions}).  In order to add the contribution of the other component,
denoted $b_L$, we replace $\hv_L$ of Eq.~(\ref{vLvR}) by $(i,1,0,0,0)^T/\sqrt{2}$
(the right-handed singlet spurion $\hv_R$ is unchanged).
Adding the two contributions together we arrive at the $SU(2)_L$ invariant
effective potential
\begin{equation}
\label{vtopeffb}
  V_{\rm eff}^{\rm top} = \frac{y^2}{2}\,C_{\rm top}
    \left( |\S_{35}-i\S_{45}|^2 + |\S_{15}+i\S_{25}|^2 \right)\ .
\end{equation}
This is the leading contribution of dynamical third-generation quarks
to the effective potential.
Expanding the non-linear $\S$ field to quadratic order in the NGB fields gives
\begin{equation}
\label{massNGBstop}
  V_{\rm eff}^{{\rm top}(2)} = 4y^2 \frac{C_{\rm top}}{f^2}\,H^\dagger H\ .
\end{equation}
When the Higgs field $H=(H_+,H_0)^T$ acquires an expectation value,
conventionally it is assigned to the lower component $H_0$, as in Eq.~(\ref{Hh}).
This selects $t_L$ (rather than any other linear combination
of the doublet fields $t_L$ and $b_L$) as the left-handed field that,
together with the right-handed field $t_R$, forms the physical top quark.

Let us pause to consider these results.
The full effective potential $V_{\rm eff}(\S)$
is the sum of Eqs.~(\ref{VeffEW}) and~(\ref{vtopeffb}).
If $C_{\rm top}$ is positive, the global minimum
is attained for $\S=\S_0={\bf 1}$ (with $\S_{i5}=0$ for $i=1,\ldots,4$).
For electroweak symmetry breaking to take place, $C_{\rm top}$
must therefore be negative.  To second order in the NBG fields,
the effective potential is the sum of Eqs.~(\ref{massNGBsEW}) and~(\ref{massNGBstop}).
As already observed in Ref.~\cite{ferretti}, the curvature at the origin
can become negative only in the direction of the $H$ field.  This happens when
\begin{equation}
\label{condition}
  \frac{C_{\rm top}}{C_{LR}}
  < -\frac{3g^2+g'^2}{2y^2}
  = -\frac{2m_W^2+m_Z^2}{m_t^2}
  \approx -0.7 \ ,
\end{equation}
triggering a non-zero expectation value for the Higgs field.

If we use Eq.~(\ref{Hh}) and, moreover, assume that all other NGB fields
in Eq.~(\ref{Pi}) remain zero, the total effective potential is
\begin{equation}
\label{Veffh}
  V_{\rm eff}(h) = -C_{LR}\left(3g^2+g'^2\right)\cos^2{(h/f)}
  + \frac{y^2}{2}\,C_{\rm top}\,\sin^2{(2h/f)}\ .
\end{equation}
While the global minimum of $V_{\rm eff}(\S)$
must occur at non-zero $h$ if Eq.~(\ref{condition}) is satisfied,
due to the complexity of $V_{\rm eff}(\S)$ we have not been able to prove that,
given arbitrary values of the SM couplings or the LECs, the global minimum
will never involve non-zero expectation values for any other NGBs.

We next turn to the calculation of the low-energy constant $C_{\rm top}$.
As in the previous subsection, this is done by matching the
effective potential~(\ref{vtopeffb}) to the underlying theory
with top-hyperbaryon couplings as given in Eq.~(\ref{lagSU5}).
The difference is that now we also integrate over the SM top quark field.
After splitting the $T_{L,R}$ spurions as in Eq.~(\ref{fsrewrite}),
the matching will involve taking four derivatives, one with respect
to each of the global spurions $v_{L,R}$ and $\bv_{L,R}$.%
\footnote{We will discuss later on what values to choose for the
  global spurions in Eq.~(\ref{fsrewrite}), or equivalently, with respect
  to which component of each global spurion one would
  choose to differentiate.
}
Once again we will set $\S={\bf 1}$.
This implies that we must take into account terms independent of $\S$
that have a similar dependence on the global spurions as Eq.~(\ref{vefftermsc}).
There are two such terms,
\begin{equation}
\label{Sindep}
  y^2 C_1(\bv_L v_L)(\bv_R v_R) + y^2 C_2(\bv_L\bv_R^T)(v_R^T v_L)\ ,
\end{equation}
where we introduced new LECs $C_1$ and $C_2$, and, for convenience,
separated out a factor of $y^2$, as we did in Eq.~(\ref{vtopeff}).
Taking the four derivatives in the effective theory we find, after
setting $\S={\bf 1}$,
\begin{equation}
\label{d4Zeff}
  \frac{\partial^4}{\partial\bv_{Li}\partial v_{Rj}\partial\bv_{Rk}
  \partial v_{L\ell}} \log Z_{\rm eff}
  =
  -y^2 V \left(C_{\rm top}\,\d_{ij}\d_{k\ell}+C_1\,\d_{i\ell}\d_{jk}
  +C_2\,\d_{ik}\d_{j\ell}\right)\ ,
\end{equation}
where $V$ is the volume.   In the microscopic theory
we find
\begin{eqnarray}
\label{d4ZUV}
  && \hspace{-10ex}
  \frac{\partial^4}{\partial\bv_{Li}\partial v_{Rj}\partial\bv_{Rk}
  \partial v_{L\ell}} \log{Z}
\\
  &=&(\l_1\l_2)^2
  \int d^4x_1\,d^4x_2\,d^4x_3\,d^4x_4
\nonumber\\
  && \times \svev{(\bB_{R\ell} t_L)(x_4)(\bt_L B_{Ri})(x_1)
  (\bB_{Lj} t_R)(x_2)(\bt_R B_{Lk})(x_3)}
\nonumber\\
  &=&-(\l_1\l_2)^2\int d^4x_1\,d^4x_2\,d^4x_3\,d^4x_4
  \int \frac{d^4p}{(2\p)^4}\frac{d^4q}{(2\p)^4}\,
       \frac{p_\m}{p^2}\,\frac{q_\n}{q^2}\, e^{ip(x_4-x_1)+iq(x_2-x_3)}
\nonumber\\
  && \times \svev{\left(\bB_{R\ell}(x_4)\g_\m P_RB_{Ri}(x_1)\right)
  \left(\bB_{Lj}(x_2)\g_\n P_LB_{Lk}(x_3)\right)}\ .
\nonumber
\end{eqnarray}
In the last equality we have integrated over the top quark,
substituting free massless fermion propagators for its two-point functions.
The remaining expectation value on the last line
is to be computed in the pure hypercolor theory.   We may now project onto the
$C_{\rm top}$ term in Eq.~(\ref{d4Zeff}) by choosing $i=j\ne k=\ell$, obtaining
\begin{eqnarray}
\label{Ctopmain}
  C_{\rm top} &=&
  \frac{(\l_1\l_2)^2}{y^2}\frac{1}{V}\int d^4x_1\,d^4x_2\,d^4x_3\,d^4x_4
  \int \frac{d^4p}{(2\p)^4}\frac{d^4q}{(2\p)^4}\,
       \frac{p_\m}{p^2}\,\frac{q_\n}{q^2}\, e^{ip(x_4-x_1)+iq(x_2-x_3)}
\nonumber\\
  && \times \svev{\left(\bB_{Rk}(x_4)\g_\m P_RB_{Ri}(x_1)\right)
  \left(\bB_{Li}(x_2)\g_\n P_LB_{Lk}(x_3)\right)}_{i\ne k}\ .
\end{eqnarray}
In terms of the Fourier transform
\begin{eqnarray}
\label{FTB}
  && \hspace{-15ex}
  \svev{\left(\bB_{R\ell}(k_4)\g_\m P_RB_{Ri}(k_1)\right)
  \left(\bB_{Lj}(k_2)\g_\n P_LB_{Lk}(k_3)\right)}
\\
  &=& \int d^4x_1\,d^4x_2\,d^4x_3\,d^4x_4\,e^{-ik_1x_1+ik_2x_2-ik_3x_3+ik_4x_4}
\nonumber\\
  && \times \svev{\left(\bB_{R\ell}(x_4)\g_\m P_RB_{Ri}(x_1)\right)
\left(\bB_{Lj}(x_2)\g_\n P_LB_{Lk}(x_3)\right)} \ ,
\nonumber
\end{eqnarray}
we may write this in momentum space as%
\footnote{In finite volume, the momentum integral $\int d^4p/(2\p)^4$
  is to be understood as a momentum average, $V^{-1} \sum_{p_\m}$.
  Alternatively, in infinite volume, $V$ is to be interpreted
  as $(2\p)^4 \d(0)$ in momentum space.
}
\begin{equation}
\label{Ctopfinal}
  C_{\rm top}
  = \frac{(\l_1\l_2)^2}{y^2}\frac{1}{V}\!
  \int\frac{d^4p}{(2\p)^4}\frac{d^4q}{(2\p)^4}\,
  \frac{p_\m}{p^2}\,\frac{q_\n}{q^2}
  \svev{\left(\bB_{Rk}(p)\g_\m P_RB_{Ri}(p)\right)
  \!\left(\bB_{Li}(q)\g_\n P_LB_{Lk}(q)\right)}_{i\ne k}\ .
\end{equation}
This is our main result.

The top sector effective potential, Eq.~(\ref{vtopeffb}),
depends on the experimentally known top Yukawa coupling, and on $C_{\rm top}$.
Using Eq.~(\ref{yukawa}), we may reexpress the ratio $\l_1\l_2/y$
in Eq.~(\ref{Ctopfinal}) in terms of quantities that are calculable
in the pure hypercolor theory.  This determines $V_{\rm eff}^{\rm top}$
completely.  The dependence on the extended hypercolor sector,
coming from the couplings $\l_{1,2}$, has dropped out.

Diagrammatically, each term on the right-hand side of Eq.~(\ref{d4Zeff})
originates from diagrams of the microscopic theory with a distinct topology.
This is shown in Fig.~\ref{topB}, where we have kept only
the propagators of the top quark (solid lines) and of the Majorana fermions
$\c$ (dashed-dot lines).  All other fields, including the Dirac fermions
in the fundamental \irrep\ of hypercolor, have been suppressed.
With these conventions, $C_{\rm top}$ arises from the class
of diagrams represented by Fig.~\ref{topB}(a), while $C_1$ and $C_2$
arise from Fig.~\ref{topB}(b) and \ref{topB}(c) respectively.

\begin{figure}[t]
\begin{center}
\includegraphics*[height=4.5cm]{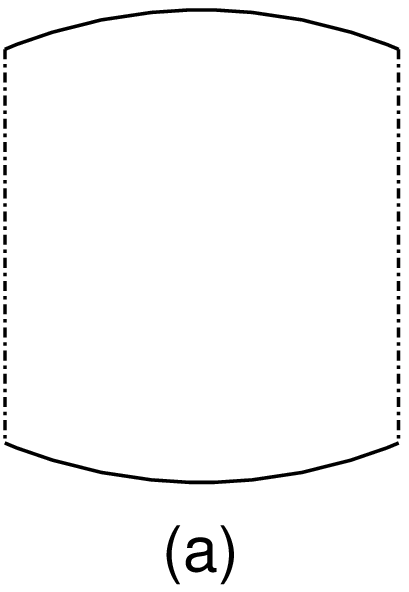}
\hspace{8ex}
\includegraphics*[height=4.5cm]{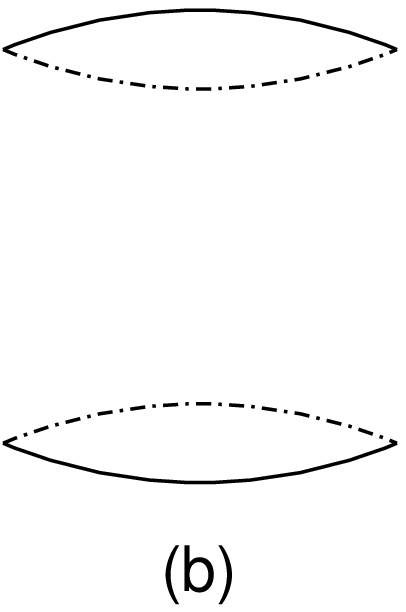}
\hspace{8ex}
\includegraphics*[height=4.5cm]{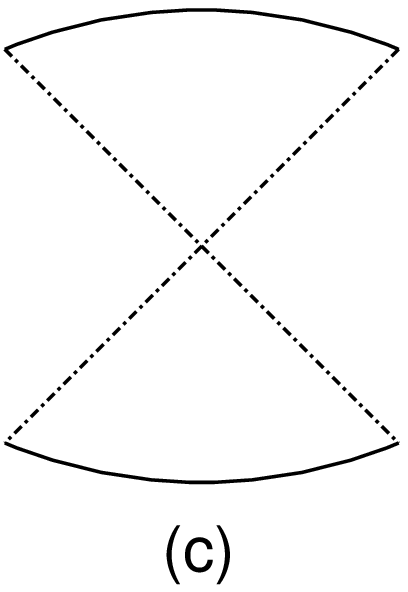}
\end{center}
\begin{quotation}
\floatcaption{topB}%
{The three possible Majorana-fermion contractions contributing to
Eq.~(\ref{d4ZUV}), corresponding to (a) $C_{\rm top}$, (b) $C_1$, and (c) $C_2$.
Only the top quark (solid lines) and Majorana fermions (dashed lines)
are shown.  The vertices $x_1,\ldots,x_4$ of Eq.~(\ref{d4ZUV})
correspond to a clock-wise motion starting at the lower-left corner.}
\end{quotation}
\vspace*{-4ex}
\end{figure}

\begin{boldmath}
\subsection{\label{large-N} Large-$N$ estimate of $y^2C_{\rm top}$}
\end{boldmath}
Determining $C_{\rm top}$ using Eq.~(\ref{Ctopfinal}) requires knowledge
of the ratio $\l_1\l_2/y$, and a strong-coupling calculation
that can be done using lattice gauge theory.
Such a lattice calculation, however, would be a major undertaking
(see Sec.~\ref{lattice}).
In this subsection, we resort to analytic techniques hoping to shed some light
on the most interesting question, which is whether $C_{\rm top}$ could
indeed be negative, and large enough in size to cause electroweak
symmetry breaking.

We will first consider what can be said if
we assume that the hyperbaryon four-point function in Eq.~(\ref{Ctopfinal})
factorizes into the product of two hyperbaryon two-point functions.
We will show that $C_{\rm top}$ is negative in this case.
We will then argue that a large-$N$ limit exists in which the factorized
contribution dominates, and thus the Higgs field acquires a non-zero
expectation value.

Assuming factorization, and using%
\footnote{On the right-hand side, $\d_{ij}$ follows from $SO(5)$ invariance.
}
\begin{eqnarray}
\label{hbprops}
  P_R\langle B_{Ri}(p)\bB_{Lj}(q)\rangle P_R
  &=& (2\p)^4 \d(p-q) \d_{ij} P_RS_{B}(p)P_R\ ,
\\
  P_L\langle B_{Li}(q)\bB_{Rj}(p)\rangle P_L
  &=& (2\p)^4 \d(p-q) \d_{ij} P_LS_{B}(p)P_L\ ,
\nonumber
\end{eqnarray}
where $S_{B}(p)$ was defined in Eqs.~(\ref{deftB}) and~(\ref{FT}),
Eq.~(\ref{Ctopfinal}) leads to
\begin{equation}
\label{Ctopfact}
  C_{\rm top}^{\rm fact}
  = -\frac{(\l_1\l_2)^2}{y^2}
  \int \frac{d^4p}{(2\p)^4}\,\frac{p_\m p_\n}{(p^2)^2}\,
  \tr\left(\g_\m P_RS_{B}(p)\g_\n P_LS_{B}(p)\right)\ .
\end{equation}
Using a dispersive representation for the hyperbaryon propagator,
\begin{equation}
\label{baryondisp}
S_{B}(p)=\int_0^\infty \frac{ds}{2\p}\,\r(s)\,\frac{-i\sl{p}+\sqrt{s}}{p^2+s}\ ,
\end{equation}
with $\r(s)\ge 0$ for all $s\ge 0$, this becomes
\begin{equation}
\label{Ctopfactdisp}
  C_{\rm top}^{\rm fact}
  = -\frac{1}{8\p^2}\,\frac{(\l_1\l_2)^2}{y^2}
  \int_0^\infty \frac{ds}{2\p}\int_0^\infty
  \frac{dt}{2\p}\,\r(s)\r(t)\,\frac{\sqrt{ts}}{t-s}\,\log{\frac{t}{s}}\ .
\end{equation}
This result is negative, because the integrand is manifestly positive.%
\footnote{Only the factorizable contribution appears to have been
  considered in Refs.~\cite{RC,ferretti}.
}
As a simple example, if we take $\r(s)=2\p f^6Z_B\d(s-M^2)$,
Eq.~(\ref{Ctopfactdisp}) reduces to
\begin{equation}
\label{Ctopfactexpl}
  C_{\rm top}^{\rm fact}
  = -\frac{1}{8\p^2}\frac{(\l_1\l_2)^2}{y^2}\,f^{12}Z_B^2
  = -\frac{1}{32\p^2}\,f^2 M^2\ ,
\end{equation}
where we used Eq.~(\ref{yukawa}).

We next consider a large-$N$ limit in which factorization
can be shown to hold.  Of course,
in the model of Ref.~\cite{ferretti}, the number of (hyper)colors is $N=4$.
What makes this generalization non-trivial is that, unlike the $SU(4)$ case
where the two-index antisymmetric \irrep\ is real, for any $N>4$ this
\irrep\ is complex.  This means that the Majorana condition~(\ref{majb})
cannot be imposed without violating gauge invariance.
In order to cope with this,
in addition to Weyl fields $\U_{ABi}$ in the antisymmetric
representation of $SU(N)$, we introduce Weyl fields $\tU_i^{AB}$
belonging to the \irrep\ made out of the antisymmetrized product
of two anti-fundamentals.%
\footnote{This is the same as the antisymmetric product of $N-2$ fundamentals,
  since $\tU^{AB} \sim \e^{A_1 A_2\dots A_{N-2} AB} \U'_{A_1 A_2\dots A_{N-2}}$.
}
Instead of Majorana fermions, we now construct Dirac fermions
out of these Weyl fields according to
\begin{eqnarray}
\label{Diracchi}
  \tom_{ABi} &=& \left(\begin{array}{c}
                \U_{ABi}\\ \e\btU_{ABi}^T
              \end{array}\right)\ ,
\\
  \bto^{AB}_i &=& \left(\begin{array}{cc}
                 -(\tU^{AB}_i)^T\e& \bU^{AB}_i
                \end{array}\right)\ ,
\nonumber
\end{eqnarray}
as well as their charge conjugates
\begin{eqnarray}
\label{defomega}
  \o^{AB}_i&=&C(\bto^{AB}_i)^T_i=\left(\begin{array}{c}
                \tU^{AB}_i\\ \e(\bU^{AB}_i)^T
              \end{array}\right)\ ,\\
  \bom_{ABi}&=&\tom^T_{ABi} C=\left(\begin{array}{cc}
                 -(\U_{ABi})^T\e& \btU_{ABi}
                \end{array}\right)\ .\nonumber
\end{eqnarray}
The index $i=1,\ldots,N_f$, now counts the number of Dirac fermions.
Going back to the $SU(4)$ theory, we have been forced to consider
an even number, $2N_f$, of Majorana fermions.  There is no
large-$N$ generalization that would involve the desired odd number
of five Majorana fermions for the $SU(4)$ theory.  Even more,
for any $N>4$ the symmetry breaking pattern becomes that of complex-\irrep\
Dirac fermions, namely, $SU(N_f)\times SU(N_f)\to SU(N_f)$ \cite{vac}.

We have shown in Fig.~\ref{topB} the contractions of the
antisymmetric-\irrep\ fermions that contribute to $C_{\rm top}$.
According to Eq.~(\ref{Ctopfinal}), we should choose fixed values $i\ne k$.
The minimal number of Dirac fermions we need in order to distinguish
Fig.~\ref{topB}(a) from Figs.~\ref{topB}(b) and~\ref{topB}(c)
is two, and so we will take
$N_f=2$ Dirac fermions in the antisymmetric \irrep.  For $N=4$,
this is equivalent to a theory with four Majorana fermions (instead of five).
This is the best we can do in terms of a large-$N$ generalization.
According to Eq.~(\ref{Ctopfinal}), for the left side of Fig.~\ref{topB}(a)
we need the contraction $P_R\svev{\o_i\bom_i}P_R$,
whereas for the right side we need $P_L\svev{\o_k\bom_k}P_L$.
Once these contractions have been fixed we can drop the indices $i$ and $k$,
and forget about the flavor index.   This suggests that the
number of flavors is not crucial if our goal is to obtain a large-$N$ estimate
of the class of diagrams depicted in Fig.~\ref{topB}(a),
and thus that the necessary transition from Majorana fermions
to Dirac fermions for $N>4$ is inconsequential.

We are now ready to give the generalization of the hyperbaryon operators.
In the $N>4$ theory with $N_f=2$ Dirac flavors they are defined by
\begin{eqnarray}
\label{cornerslN}
  B_{Ria} &=& \e_{abc}\,P_R\o_i^{AB}\,(\J_{Ab}^T\,\e\,\J_{Bc})\ ,
\\
  \bB_{Ria} &=& \e_{abc}\,\bom_{iAB}P_L\,\Big(\bJ^{Ab}\,\e\,(\bJ^{Bc})^T\Big)\ ,
\nonumber\\
  B_{Lia} &=& \e_{abc}\,P_L\o_i^{AB}\,(\J_{Ab}^T\,\e\,\J_{Bc})\ ,
\nonumber\\
  \bB_{Lia} &=& \e_{abc}\,\bom_{iAB}P_R\,\Big(\bJ^{Ab}\,\e\,(\bJ^{Bc})^T\Big)\ .
\nonumber
\end{eqnarray}
We have used index conventions similar to the previous sections.
For $N=4$ we may impose the Majorana condition $\o=\tom$,
and then these definitions reproduce Eq.~(\ref{bar4}).

\begin{figure}[t]
\begin{center}
\includegraphics*[height=5cm]{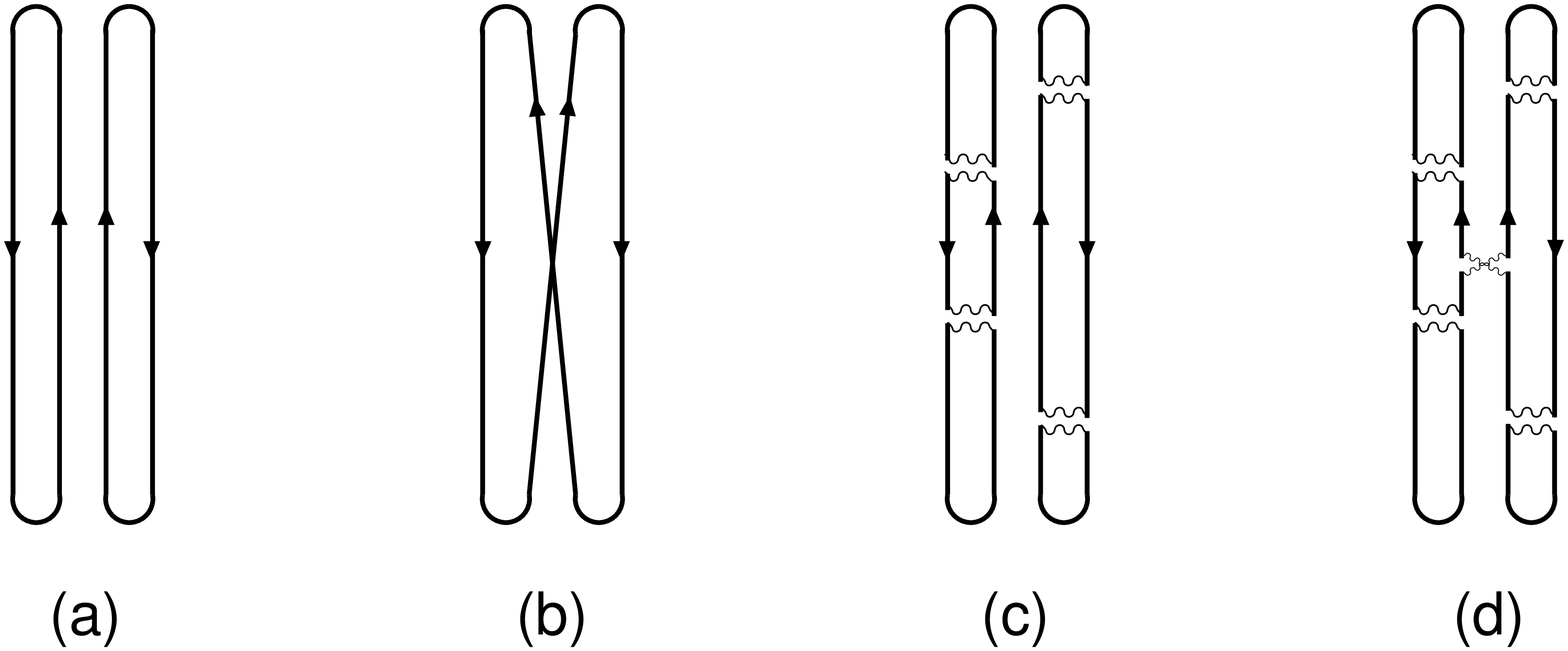}
\end{center}
\begin{quotation}
\floatcaption{baryonN}%
{A hyperbaryon in large $N$: diagrams without hypergluons (a,b)
and with them (c,d).  Diagrams (a) and (c) are leading, while
(b) and (d) are subleading.}
\end{quotation}
\vspace*{-4ex}
\end{figure}

Before we work out the more complicated case of Fig.~\ref{topB}(a),
let us consider the behavior of a single hyperbaryon in large $N$.
The hyperbaryons are bound by interchanging hypergluons between their
elementary constituents.  The situation here is different from the conventional
large-$N$ limit of baryons made only of fundamental-\irrep\ fermions,
where the number of constituents grows linearly with $N$ \cite{WLN}.
For our hyperbaryons, the number of elementary constituents (as well as
the number of hypercolor indices of each field) is fixed.
This resembles the behavior of mesons within the usual large-$N$ treatment.

In Fig.~\ref{baryonN} we show a few examples.  The bottom end of each diagram
represents a hyperbaryon, say $B_R$ (first line of Eq.~(\ref{cornerslN})),
and the top end the corresponding anti-baryon,
say $\bB_L$ (last line of Eq.~(\ref{cornerslN})).
Starting with Fig.~\ref{baryonN}(a)
the two vertical lines in the middle represent the propagation
of the double-indexed fermion of the antisymmetric \irrep,
and the lines on the sides
the propagation of the two fundamental-\irrep\ fermions, one on each side.
The lines are oriented: the arrows point from a superscript index to a
subscript index.  Fig.~\ref{baryonN}(b) shows
an alternative index contraction, still without hypergluon fields.
Note that Fig.~\ref{baryonN}(a) dominates over Fig.~\ref{baryonN}(b)
in large $N$, since the former is of order $N^2$ and the latter of
order $N$.  In Fig.~\ref{baryonN}(c) we have added hypergluon interactions.
Introducing the `t~Hooft coupling $\l=g^2 N$ it follows that
that planar diagram is again of order $N^2$.  The factors of $g=\sqrt{\l/N}$
from each hypergluon vertex are compensated by
a matching increase in the number of index loops.
While the diagram appears disconnected
to the eye, this is really not the case, because the two central vertical
lines correspond to the two-point function of a single two-index fermion,
$\svev{\o^{AB}\bom_{CD}}$.
Fig.~\ref{baryonN}(d) shows a different arrangement of hypergluon
interactions.  The hypergluon that is exchanged at the center of the diagram
represents a self-energy correction for $\svev{\o^{AB}\bom_{CD}}$,
which, to be consistent with the directionality of the index lines,
gives rise to a non-planar diagram.  This diagram is subleading
in the large-$N$ counting.

The upshot is that,
in large $N$, the dominant diagrams that bind the hyperbaryon
are planar diagrams of order $N^2$, such as for example those in
Fig.~\ref{baryonN}(a) and Fig.~\ref{baryonN}(c).
The hyperbaryon two-point function in Eq.~(\ref{FT}) will exhibit this large-$N$
behavior, much like the two-point function of NGBs made out of
antisymmetric-\irrep\ fermions \cite{taco}, which, in turn,
leads to $f\sim N$ for large $N$.%
\footnote{Note the difference with QCD, where the decay constant $f_\p$
  of the fundamental-\irrep\ NGBs scales like $\sqrt{N}$.
}
Using Eqs.~(\ref{B}) and~(\ref{yukawa}),
it follows that $M$ is independent of $N$, while $Z_B\sim 1/N^4$,
and the top Yukawa coupling behaves like $y\sim N$.
In contrast to QCD, where $m_{\rm nucleon}/f_\p$ grows like $\sqrt{N}$,
we find that $M/f$ decreases like $1/N$.
If $M$ is indeed related to the mass of the lightest hyperbaryon in the theory,
this suggests that the lightest hyperbaryon could be relatively light
compared to $\L_{\rm HC}$.

With the contractions of the double-index hyperfermions fixed to be those in
Fig.~\ref{topB}(a), let us now study the large-$N$ behavior of the various
possible contractions of the single-index, fundamental-\irrep\ fields.
Since $U(1)_X$ is not broken spontaneously, the Wick contractions
have to comply with this symmetry.
Let us start, for example, with the two $\J$ fields of the $B_R$
hyperbaryon at the bottom left of Fig.~\ref{topB}(a).
They can be contracted with the corresponding
fields in the top left or the bottom right, but not with those in the top right.
There are three possibilities:  (1) both $\J$ fields are contracted with those
at the top left, (2) both are contracted with those at the bottom right,
or (3) one is contracted with a field at the top left and the other
with a field at the bottom right.

\begin{figure}[t]
\begin{center}
\includegraphics*[height=5cm]{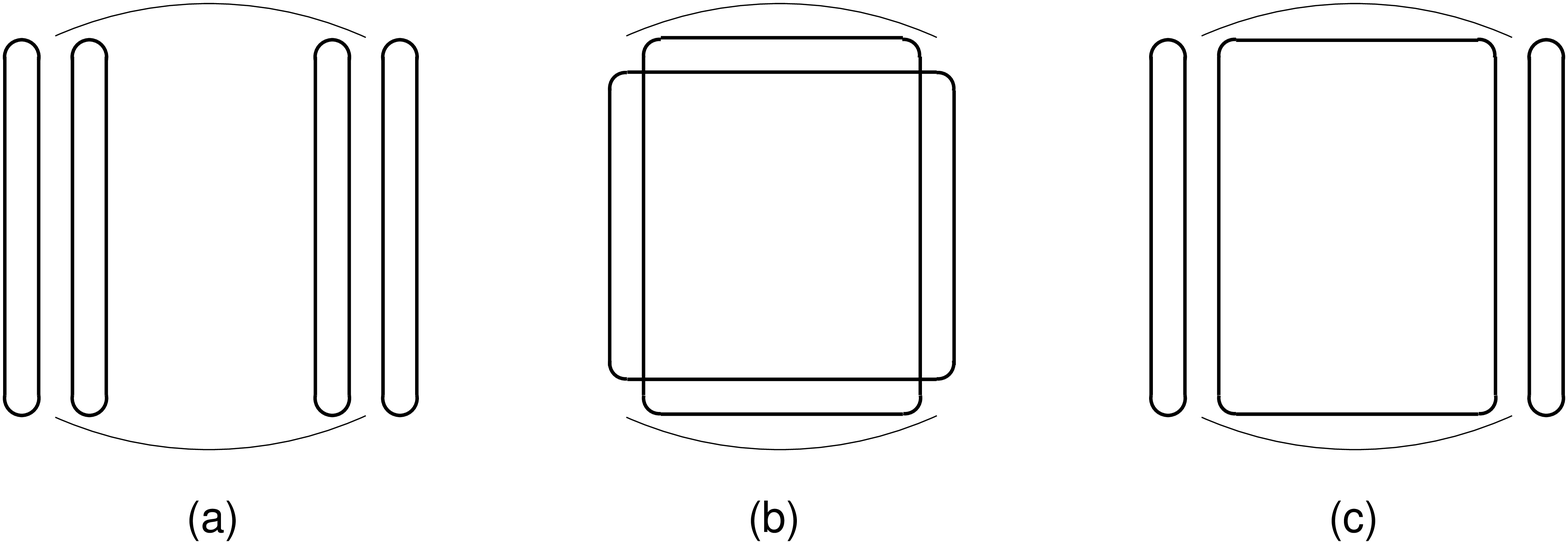}
\end{center}
\begin{quotation}
\floatcaption{ctopN}%
{Different large-$N$ contributions to Fig.~\ref{topB}(a).
The curved lines at the top and the bottom represent the two top-quark
propagators.  The diagrams scale like (a) $N^4$, (b) $N^2$, and (c) $N^3$}
\end{quotation}
\vspace*{-4ex}
\end{figure}

Let us consider these three cases in turn.
First we contract both $\J$ fields at the lower left corner
with the $\bJ$ fields of the $\bB_L$ hyperbaryon
at the upper left corner, \ie, case (1) above.
Remembering that also the two-index field at the lower left corner of
Fig.~\ref{topB}(a) is contracted with the two-index field at the
upper left corner, this gives a contribution proportional to
\begin{equation}
\label{contr1}
  \left(\d^A_C\d^B_D-\d^A_D\d^B_C\right)
  \left(\d_A^C\d_B^D-\d_A^D\d_B^C\right)=2N(N-1)\ ,
\end{equation}
where we label the $SU(N)$ indices as $\o^{AB}\J_A\J_B$ at the lower left corner
and as $\bom_{CD}\bJ^C\bJ^D$ at the upper left corner.   We get a
similar factor from the right side of the diagram, so that the total diagram
is of order $(N(N-1))^2\sim N^4$.  In Fig.~\ref{ctopN}(a)
we show as an example the order-$N^4$ contribution coming
from using twice the diagram of Fig.~\ref{baryonN}(a).

For case (2), the $\J$ fields at the lower left corner are both contracted
with the corresponding fields at the lower right corner.
Using $SU(N)$ indices $AB$ at the lower left,
$CD$ at the upper left, $EF$ at the lower right, and $GH$ at the upper
right corners, this contraction leads to a contribution of order
\begin{equation}
\label{contr2}
  \d^{[A}_C\d^{B]}_D\d^{[G}_E\d^{H]}_F\d^{[E}_A\d^{F]}_B\d^{[C}_G\d^{D]}_H\sim N^2\ .
\end{equation}
The notation $[\dots]$ denotes antisymmetrization in the pair of indices inside
the brackets.  An example is shown in Fig.~\ref{ctopN}(b).

Finally we consider the mixed case (3), where one of the $\J$ fields is
contracted with the lower right corner, and the other with the upper left one.
It is straightforward to see that this gives a contribution
\begin{equation}
\label{contr3}
  \d^{[A}_C\d^{B]}_D\d^{[G}_E\d^{H]}_F\d_A^{[C}\d_G^{D]}\d_B^{[E}\d_H^{F]} \sim N^3\ .
\end{equation}
An example is shown in Fig.~\ref{ctopN}(c).

We should also consider diagrams ``dressed''
with hypergluons.   The interesting case is that of Fig.~\ref{ctopN}(a),
where hypergluons exchanged between the two hyperbaryons on the
left and on the right make these hyperbaryons interact.   Such interactions
would spoil the factorization of the four-point function in Eq.~(\ref{Ctopmain}).
However, these interactions are suppressed in large $N$ for the same
reason that meson-meson interactions are suppressed in large-$N$ QCD.
Any hypergluon connecting the left and right sides of Fig.~\ref{ctopN}(a)
will reduce the number of hypercolor loops by one,
in addition to adding a factor of $g^2/N$.
The key point here is that the number of hyperfermion
constituents of the hyperbaryons is fixed to three, in contrast to the
case of baryons in QCD, where the number of constituent quarks grows
linearly with $N$.

We conclude that the factorizable part of Fig~\ref{topB}(a),
which grows like $N^4$, is the dominant large-$N$ contribution of the
top-quark sector.  $C_{\rm top}$ itself grows like $N^2$,
since $V_{\rm eff}^{\rm top}$ is proportional to $y^2 C_{\rm top}$
and $y$ scales like $N$ (this is consistent with the large-$N$ behavior
of the right-hand side of Eq.~(\ref{Ctopfactexpl})).
According to the discussion in the beginning of this subsection,
this produces a negative
value for $C_{\rm top}$, which, in turn, generates a negative curvature
for the Higgs field in Eq.~(\ref{Veffh}).

We recall that the other contribution to the effective potential~(\ref{Veffh})
for the Higgs field is coming from a single electroweak gauge-boson exchange.
This contribution, which is parametrized by $C_{LR}$, always produces
a positive-curvature term in the effective potential.
It is easily seen that $C_{LR}$ is subleading in the large-$N$ counting.
The electroweak gauge group $SU(2)_L$ is a subgroup of $SU(5)$,
whereas hypercharge is a subgroup of $SU(5)\times U(1)_X$.
Therefore all the electroweak gauge bosons interact with the
antisymmetric-\irrep\ hyperfermions, from which the $SU(5)/SO(5)$ coset fields
are formed.  $C_{LR}$ involves a single closed fermion loop,
and so for the antisymmetric-\irrep\ fields it is of order $N^2$.
(For the fundamental \irrep, $C_{LR}$ would be of order $N$.
  Observe that the electroweak gauge bosons do not carry hypercolor,
  and therefore their exchange has no effect on the large-$N$ counting.)
This is subleading to the factorizable contribution of the top sector,
which is of order $N^4$.
It follows that, in large $N$, Eq.~(\ref{Veffh}) is dominated
by the contribution of the top-quark sector, developing a negative curvature
at the origin that triggers electroweak symmetry breaking.

\bigskip

\subsection{\label{pheno} Phenomenological consequences}
Our analysis in this section was based on the lagrangian~(\ref{lagSU5}).
While $\cl_{\rm EHC}$ depends on two coupling constants $\l_1$ and $\l_2$,
only their product enters the determination of the top Yukawa coupling
in Sec.~\ref{top-mass}, and of the induced effective potential for the Higgs
studied in Sec.~\ref{top-effpot}.  An additional free parameter of the model
is the scale of the hypercolor theory itself, $\L_{HC}$.
Using the results of Sec.~\ref{top-mass} and Sec.~\ref{top-effpot},
we may in principle perform a lattice calculation that will
fix the values of $\l_1\l_2$ and of $\L_{HC}$
in terms of the experimentally known values of the top Yukawa coupling $y$,
and of the Higgs field's expectation value,
provided that $C_{\rm top}$ turns out to be large enough compared
to $C_{LR}$ to trigger condensation (see Eq.~(\ref{condition})).%
\footnote{We are assuming that none of the $SU(2)_L$ triplet NGBs acquires
  an expectation value.
}

The only remaining uncertainty then arises from the four different choices
for the hyperbaryon fields (that each can be a primed or an unprimed one,
\seef\ Table~\ref{tabHC})
in Eq.~(\ref{lagSU5}).  This gives rise to a discrete four-fold ambiguity
in the predicted values of $\l_1\l_2$ and of $\L_{HC}$.
For each of these four possibilities, one can proceed to
compare other predictions of the hypercolor theory
with experimental constraints, which can in principle rule out, or rule in,
that particular version of the hypercolor theory.

A less constrained hypercolor model can be obtained by relaxing the assumption
that $SU(3)\times SU(3)'$ is broken explicitly to $SU(3)_{\rm color}$ only by
the QCD interactions.  There are at least two alternative ways
to introduce such an explicit breaking.

First, one can introduce a Dirac mass term $m\bj\j$ for the fundamental-\irrep\
hyperfermions, with $m\geqx\L_{HC}$.  We may think of $m$ as arising
from the expectation value of a three by three global spurion $M_{ab}=m\d_{ab}$,
where $M_{ab}$ has the same transformation properties as the
$SU(3)\times SU(3)'/SU(3)_{\rm color}$ coset field $\O_{ab}$.
Using the global spurion $M_{ab}$ allows for the coupling of each SM spurion
field $T_{L,R}$ to both of the unprimed an primed hyperbaryon fields
from Table~\ref{tabHC}.  In fact,
if the hyperbaryon field is assumed to have well defined
transformation properties under $SU(3)_{\rm color}$ only,
some additional operators besides those shown in Table~\ref{tabHC}
may occur.  For their structure, see Table~2 of Ref.~\cite{ferretti}.
The bottom line is that $\cl_{\rm EHC}$ can now depend on four (or more)
coupling constants, instead of just two.

An alternative mechanism involves the introduction of two more spurions
$T'_{L,R}$, and assuming that the unprimed hyperbaryon fields and spurions
transform under $SU(3)$, while the primed ones transform under $SU(3)'$.
Twice as many terms are then allowed by $SU(3)\times SU(3)'$ invariance
of $\cl_{\rm EHC}$, with each term involving either unprimed or primed fields.
The explicit breaking to $SU(3)_{\rm color}$
then occurs by assigning to all spurions, both primed and unprimed,
the same SM values as in Eq.~(\ref{fspurions}).

In any one of these more general schemes, the matching
of the top Yukawa coupling and of $C_{\rm top}$ to the underlying theory
can be done using the same techniques as before.
However, the predictive power will be reduced, since this analysis
would still provide just two constraints on the larger set of parameters,
which includes $\L_{HC}$ and all the coupling constants that may now occur
in $\cl_{\rm EHC}$.

Finally, note that if we expand both $\S$ and $\F$ in Eq.~(\ref{topmasseff})
to first order in the NGB fields, and use the SM values $\hT_{L,R}$,
we obtain the dimension-five operator
$(y/f') \eta'\bt_R H_i \e_{ij} q_{Lj}$ and its hermitian conjugate,
where $f'$ is the decay constant of $\eta'$.  This couples $\eta'$
to the SM fields.  If we set the Higgs field to its vacuum expectation value,
the above operator reduces to $(m_t/f') \eta' \bt_R t_L$.
The phenomenological implications of these interactions
have to be looked into.  If they turn out to be incompatible with experiment,
this would necessitate the introduction of an explicit breaking of $U(1)'$
that makes the $\eta'$ sufficiently heavy.  One possible source for this
explicit breaking is the Dirac mass term $m\bj\j$ discussed above.

\section{\label{lattice} Lattice aspects}
In this short section, we explain why a lattice computation of $C_{\rm top}$
would be a ``QCD-like'' computation.  The question to address
is why the lattice formulation of hypercolor theories such as
considered here resembles the lattice formulation of QCD, in view of the
well-known complications with fermion doubling and
chirality on the lattice.   The fermion doubling problem has its roots
in the observation that a single Weyl fermion cannot live on the lattice
\cite{KS,NN}.  Since the hypercolor theory contains an odd number of Weyl
fermions in the two-index antisymmetric \irrep,
this might seem to imply that the model we consider
here cannot be easily discretized.

The first observation is that the integration over the SM quark fields
$q_L$ and $t_R$ is done analytically, as we did in this article.
The results for the top Yukawa coupling $y$ (Eqs.~(\ref{mu}) and~(\ref{y}))
and for $C_{\rm top}$ (Eq.~(\ref{Ctopfinal})) are obtained in terms of
pure hypercolor correlation functions.   Hence only the hypercolor theory
needs to be considered on the lattice, and the lattice action
does not contain the four-fermion lagrangian $\cl_{\rm EHC}$ of Eq.~(\ref{lagSU5}).

Let us start from the sector that resembles QCD most closely, namely, the
Dirac fermions $\j_a$ in the fundamental \irrep\ of $SU(4)$ hypercolor.
These can be treated in exactly the same way as the quark fields of
$N_f$-flavor QCD.
In the Wilson formulation of the theory, the gauge invariant Wilson mass term
removes the fermion doublers at the price of breaking the symmetry group
$SU(N_f)_L\times SU(N_f)_R$ of the continuum theory explicitly to
its diagonal $SU(N_f)$ subgroup.
Tuning the bare mass appropriately, one then recovers the massless
theory with the full chiral symmetry group in the continuum limit \cite{KS}.
In the hypercolor theory we have $N_f=3$ Dirac fermions.
The chiral flavor group $SU(3)_L\times SU(3)_R$ has been renamed
$SU(3)\times SU(3)'$, and the unbroken diagonal subgroup was identified
with $SU(3)_{\rm color}$.

We next turn to the novel feature of the hypercolor theory, which
is the 5 real-\irrep\ Weyl fermions $\U_{ABi}$.
The key point is that each of these Weyl fermions can be assembled
together with its antifermion into a Majorana fermion field $\c_{ABi}$.
A Majorana mass term of the form $\bc^{AB}_i\c_{ABi}$ is allowed by the $SU(4)$
gauge symmetry, just like a Dirac mass term is allowed in the familiar case
of fundamental-\irrep\ fermions.
Therefore, a Wilson formulation of the Majorana fermions
is possible, with a Wilson mass term that once again removes
fermion doublers without breaking gauge invariance.
The hypercolor theory is chiral with respect to the $SU(5)$ flavor symmetry
of the Majorana or Weyl fields, and consequently,
the Majorana--Wilson mass term breaks $SU(5)$ explicitly down to $SO(5)$.
Again, the full $SU(5)$ chiral symmetry
should be recovered in the continuum limit after appropriate tuning
of the bare mass.%
\footnote{The lattice formulation of an adjoint Majorana fermion
  was studied extensively in the context of supersymmetric theories,
  see, \eg, Ref.~\cite{susy}.
}

The situation with respect to (complex-\irrep)
Dirac fermions and to (real-\irrep) Majorana fermions is thus
completely parallel.  In both cases,
what we anticipate as the spontaneous symmetry breaking pattern
of the continuum theory turns into an explicit breaking in the Wilson
formulation.  The explicit breaking disappears, and the full chiral
symmetry group can be recovered, in the continuum limit by tuning
the bare mass terms.  In particular, the flavor group $SO(5)$
of the Majorana-Wilson fermion action in the hypercolor theory
enlarges to $SU(5)$ in the continuum limit, much like the
diagonal $SU(N_f)$ symmetry of the Dirac-Wilson action of QCD enlarges
to the full $SU(N_f)_L\times SU(N_f)_R$ symmetry in the continuum limit.

In short, it is precisely the unbroken flavor symmetry group $H$
of Eq.~(\ref{breaking}) that is preserved in a lattice formulation
with Wilson fermions.   It is actually possible
to gauge the SM group $SU(3)_{\rm color}\times SU(2)_L\times U(1)_Y$
in this lattice formulation, because this group is contained in $H$.
The difficulties caused on the lattice with chiral gauge symmetries%
\footnote{For a review, see \cite{MGlat2000}.}
would only appear were one to couple also the SM fermions to this model.

The lattice formulation of the hypercolor theory is not without technical
challenges.  First, two fermion \irreps\ need to be introduced simultaneously.
While clearly a coding task, this is
something that to our knowledge has not been done to date.  Also,
the computation of a four-point function as in Eq.~(\ref{Ctopmain})
would be very demanding.  Because there are two independent momentum variables,
the cost is expected to grow like the square of the 4-volume of the lattice.
The computation of $S_{B}(0)$ in Eq.~(\ref{mu}),
and of the factorizable contribution to $C_{\rm top}$, would already be
interesting.   Here only the hyperbaryon two-point function is required,
and the cost grows linearly with the 4-volume.

\section{\label{conclusion} Conclusion}
In this article, we discussed a recently proposed composite Higgs model
\cite{ferretti}, concentrating on the top quark sector.
This ``hypercolor'' model is an
$SU(4)$ gauge theory, with a quintuplet of two-index antisymmetric
Majorana fermions and a color triplet of $SU(4)$-fundamental Dirac fermions.
We considered the
top Yukawa coupling and the top contribution to the effective potential
for the Higgs, showing how to match the relevant low-energy constants
$\m=yf/2$ and $C_{\rm top}$ to correlation functions of the hypercolor theory.

This matching is the starting point for any non-perturbative evaluation
of the low-energy constants.   The needed computations can in principle
be done on the lattice, using methods that are much the same as those
employed in lattice QCD.   The main differences are that now the gauge
group is $SU(4)$ instead of $SU(3)$, and that there are fermions in
more than one \irrep\ of the gauge group.
For the Higgs effective potential, a four-point hyperbaryon correlation
function needs to be considered; this appears to have been overlooked in
at least some of the literature on models of this type.  A lattice calculation
of this four-point function would be very demanding, and how to do it
efficiently is a question that goes beyond the scope of this paper.
As mentioned in Sec.~\ref{lattice}, only a two-point function is needed
for the low-energy constant $\mu$,
as well as for the factorizable contribution to $C_{\rm top}$.
This computation would be comparable in scope with a lattice computation
of $C_{LR}$, the LEC controlling the
contribution to the Higgs effective potential from
the SM gauge bosons (\seef\ Sec.~\ref{weak-effpot}).

We found that a large-$N$ limit exists in which the factorizable
contribution to the hyperbaryon four-point function dominates $C_{\rm top}$.
We also showed that the factorizable contribution generates a negative
curvature at the origin.  This is a necessary condition for electroweak
symmetry breaking, because the effective potential induced by
the SM gauge bosons does not break the SM symmetries,
a manifestation of vacuum alignment.

For very large $N$, the top sector dominates the whole
Higgs effective potential.  This maximizes the symmetry breaking,
with the minimum of $V_{\rm eff}(h)$ presumably occurring for $\sin(2h/f)=1$
in Eq.~(\ref{Veffh}).   Phenomenologically, this is not allowed \cite{RC,ferretti}.
The hope is that for $N=4$, electroweak symmetry breaking
with a phenomenologically acceptable value of $h/f$ takes place.
Only the lattice can address this question quantitatively.

In order to couple the SM fermions to the hypercolor theory,
an extended hypercolor sector is necessary.  While the detailed structure
of the EHC theory is very important for phenomenology,
the lattice setup is largely blind to these details.  At energy scales
much below $\L_{\rm EHC}$,
the coupling of SM and hypercolor fermions is summarized by
the four-fermion lagrangian $\cl_{\rm EHC}$ of Eq.~(\ref{lagSU5}).
Much like the familiar treatment of hadronic matrix elements
of the electroweak interactions,
$\cl_{\rm EHC}$ is not taken as part of the lattice action.
Instead, one evaluates the hypercolor-theory correlation functions
that arise from working to leading order in the four-fermion lagrangian
$\cl_{\rm EHC}$.

In the most constrained case,
which is the one we have worked out in detail in this paper,
$\cl_{\rm EHC}$ depends on only two couplings $\l_{1,2}$.
A lattice computation can then in principle determine
their product $\l_1\l_2$ together with the hypercolor scale $\l_{\rm HC}$
in terms of the experimental values of the top Yukawa coupling
and the Higgs expectation value, up to a four-fold ambiguity.
In a less constrained setup, a similar lattice computation would supply two
constraints among the parameters of the hypercolor theory.

The parameters $\l_{1,2}$ have mass dimension two.
According to naive dimensional analysis, their values would be
of order $(\L_{\rm HC}/\L_{\rm EHC})^2$, making the top Yukawa coupling
of order $y \sim (\L_{\rm HC}/\L_{\rm EHC})^4$.  For comparison,
we recall that in classic (walking) technicolor,
the top Yukawa coupling is naively of order $(\L_{\rm TC}/\L_{\rm ETC})^2$,
where $\L_{\rm TC}$ and $\L_{\rm ETC}$ are the scales of the technicolor
and of the extended technicolor theories.  Naively, the case
for a partially-composite top seems even worse than for technicolor.
On the other hand,
it might be that the experimental constraints allow $\L_{\rm EHC}$ to be
much closer to $\L_{\rm HC}$ than $\L_{\rm ETC}$ to $\L_{\rm TC}$.
We also found that the hyperbaryons that couple to the top quark
in $\cl_{\rm EHC}$ might be relatively light, as suggested by large-$N$ counting,
and this helps boost the value of the top Yukawa $y$ as well.  The
key reason for the possible lightness of these hyperbaryons is that
they are composed of hyperfermions in two different \irreps\ of the
gauge group.  As far as large-$N$ counting goes, these hyperbaryons behave
more like mesons than like baryons in the usual large-$N$ limit of QCD.

Yet another feature that may be needed for a phenomenologically viable
partial-compositeness model is large anomalous dimensions
for the hyperbaryon fields \cite{RC}.  It should be possible to
study on the lattice whether or not this is the case.

\vspace{3ex}
\noindent {\bf Acknowledgments}
\vspace{3ex}

We thank Gabriele Ferretti for explanations about phenomenological
aspects of the model, Will Jay for a discussion of hyperbaryon operators,
and Tom DeGrand for his comments on the manuscript.
We also thank Ben Svetitsky, Yuzhi Liu and Ethan Neil
for discussions.  MG and YS thank the Department of Physics of
the University of Colorado, Boulder, and
YS thanks the Department of Physics and Astronomy of San Francisco
State University for hospitality.
MG is supported in part by the US Department of Energy, and
YS is supported by the Israel Science Foundation
under grant no.~449/13.

\appendix
\section{\label{conv} Conventions}
We choose
our $\g$ matrices
to be hermitian, and
we use the chiral representation
\begin{equation}
  \g_k =
  \left(
  \begin{array}{cc}
    0      & i\s_k  \\
    -i\s_k & 0
  \end{array}
  \right)
\,,\qquad
  \g_4 =
  \left(
  \begin{array}{cc}
    0  &  1  \\
    1  &  0
  \end{array}
  \right) \,,
\label{4dDirac}
\end{equation}
with $\s_k,$ $k=1,2,3,$ the Pauli matrices.  The chiral projectors are
$P_R = (1+\g_5)/2$, $P_L = (1-\g_5)/2$, where
\begin{equation}
  \g_5 = -\g_1\g_2\g_3\g_4
  = \left(
  \begin{array}{cc}
    1  &  0  \\
    0  & -1
  \end{array}
  \right) \,.
\label{pm5}
\end{equation}
The charge conjugation matrix occurring in Eq.~(\ref{maj})
is $C=-\g_2\g_4$.  It satisfies
\begin{equation}
  C \g_\m = - \g_\m^T C \,,
\label{C}
\end{equation}
and $C^{-1}=C^\dagger=C^T=-C$.

For the invariant $SU(2)$ subgroups of $SO(4)$ we may choose
the generators as the following tensor products of Pauli matrices
and the $2\times 2$ identity matrix $I$,
\begin{eqnarray}
  2\, T_L^1&=&  \sigma_2 \times \sigma_1 \ ,
\label{tensorprod}\\
  2\, T_L^2 &=& -\sigma_2 \times \sigma_3 \ ,
\nonumber\\
  2\, T_L^3 &=& I \times \sigma_2 \ ,
\nonumber\\
  2\, T_R^1 &=& \sigma_1 \times \sigma_2 \ ,
\nonumber\\
  2\, T_R^2 &=& \sigma_2 \times I \ ,
\nonumber\\
  2\, T_R^3 &=& \sigma_3 \times \sigma_2 \ .
\nonumber
\label{tensor}
\end{eqnarray}
Identifying $SO(4)$ with the upper-left $4\times4$ block,
the $SU(2)_L$ generators and the third  $SU(2)_R$ generator
(which is used to construct the hypercharge $Y$) are given explicitly by
\begin{eqnarray}
\label{embed}
T_L^1&=&\frac{i}{2}\left(\begin{array}{ccccc}
0 & 0 & 0 & -1 & 0 \\
0 & 0 & -1 & 0 & 0 \\
0 & 1 & 0 & 0 & 0 \\
1 & 0 & 0 & 0 & 0 \\
0 & 0 & 0 & 0 & 0\end{array}\right)\ ,\qquad
T_L^2 \ = \ \frac{i}{2}\left(\begin{array}{ccccc}
0 & 0 & 1 & 0 & 0 \\
0 & 0 & 0 & -1 & 0 \\
-1 & 0 & 0 & 0 & 0 \\
0 & 1 & 0 & 0 & 0 \\
0 & 0 & 0 & 0 & 0\end{array}\right)\ , \hspace{5ex}
\\
T_L^3&=& \rule{0ex}{11ex}
\frac{i}{2}\left(\begin{array}{ccccc}
0 & -1 & 0 & 0 & 0 \\
1 & 0 & 0 & 0 & 0 \\
0 & 0 & 0 & -1 & 0 \\
0 & 0 & 1 & 0 & 0 \\
0 & 0 & 0 & 0 & 0\end{array}\right)\ ,\qquad
T_R^3 \ = \ \frac{i}{2}\left(\begin{array}{ccccc}
0 & -1 & 0 & 0 & 0 \\
1 & 0 & 0 & 0 & 0 \\
0 & 0 & 0 & 1 & 0 \\
0 & 0 & -1 & 0 & 0 \\
0 & 0 & 0 & 0 & 0\end{array}\right)\ .\nonumber
\end{eqnarray}

\vspace{5ex}

\end{document}